\documentclass[a4paper,11pt]{article}
\usepackage{ifpdf} 
\ifpdf
\pdfoutput=1 
\fi

\usepackage{jcappub}
\usepackage{graphicx}
\graphicspath{{Paper/}{./}}
\usepackage{dcolumn}
\usepackage{amssymb,amsmath,bm,bbold}
\usepackage{color}
\usepackage[dvipsnames]{xcolor}
\usepackage{xfrac}
\usepackage{aas_macros}
\usepackage{mathrsfs}
\usepackage{subcaption}
\usepackage{rotating}
\usepackage{chngcntr}

\newcommand{\vd}{\mathbf{d}}
\newcommand{\vt}{\mathbf{t}}

\newcommand{\vN}{\mathbf{N}}

\newcommand{\sC}{{\sf C}}

\definecolor{internationalkleinblue}{rgb}{0.0, 0.18, 0.65}
\hypersetup{urlcolor=internationalkleinblue, linkcolor=internationalkleinblue, citecolor=internationalkleinblue}

\usepackage[T1]{fontenc} 
\usepackage{natbib}
\title{Analytic marginalization of $N(z)$ uncertainties in tomographic galaxy surveys}

\author[a,1]{Boryana Hadzhiyska,}
\author[b]{David Alonso,}
\author[c]{Andrina Nicola,}
\author[d]{An\v{z}e Slosar}

\affiliation[a]{Harvard-Smithsonian Center for Astrophysics, 60 Garden St., Cambridge, MA 02138, USA}
\affiliation[b]{Department of Physics, University of Oxford, Denys Wilkinson Building, Keble Road, Oxford OX1 3RH, United Kingdom}
\affiliation[c]{Department of Astrophysical Sciences, Princeton University, Peyton Hall, Princeton NJ 08544-0010, USA}
\affiliation[d]{Brookhaven National Laboratory, Physics Department, Upton, NY 11973, USA}
\emailAdd{boryana.hadzhiyska@cfa.harvard.edu}

\abstract{We present a new method to marginalize over uncertainties in redshift distributions, $N(z)$, within tomographic cosmological analyses applicable to current and upcoming photometric galaxy surveys. We allow for arbitrary deviations from the best-guess $N(z)$ governed by a general covariance matrix describing the uncertainty in our knowledge of redshift distributions. In principle, this is marginalization over hundreds or thousands of new parameters describing potential deviations as a function of redshift and tomographic bin. However, by linearly expanding the theory predictions around a fiducial model, this marginalization can be performed analytically, resulting in a modified data covariance matrix that effectively downweights the modes of the data vector that are more sensitive to redshift distribution variations. We showcase this method by applying it to the galaxy clustering measurements from the Hyper Suprime-Cam first data release. We illustrate how to marginalize over sample-variance of the calibration sample and a large general systematic uncertainty in photometric estimation methods, and explore the impact of priors imposing smoothness in the redshift distributions.}

\begin{document}
\maketitle
\flushbottom

  \section{Introduction}\label{sec:intro}
    Photometric galaxy surveys  have the potential to transform our understanding of the Universe by measuring the properties of millions and soon billions of galaxies on the sky. These catalogs can be used to constrain cosmological parameters from measurements of galaxy clustering and weak gravitational lensing as demonstrated by numerous ongoing surveys, including the Sloan Digital Sky Survey \cite{astro-ph/0006396,2013MNRAS.432.1544M}, PanSTARRS \cite{2010SPIE.7733E..0EK}, Kilo-degree Survey \cite{1206.1254,2020A&A...633A..69H}, Dark Energy Survey \cite{1601.00329,2018PhRvD..98d3526A}, Hyper Suprime-Cam survey (HSC) \cite{2012SPIE.8446E..0ZM,2019PASJ...71...43H,1912.08209}. These cosmological probes will also play a crucial role in the upcoming Vera Rubin Observatory Legacy Survey of Space and Time \cite{0912.0201}.

    It has been long recognized that the accuracy of cosmological  constraints  derived from photometric surveys relies heavily on the accuracy of photometric redshift distributions and that this is likely to continue to be the dominant source of systematic uncertainty. The crucial quantity is the redshift distribution $N(z)$, which is the mean number of galaxies as a function of redshift for each tomographic sample \cite{1809.01669}.  Improving photometric redshift techniques and calibration methods for redshift distributions is an area of active research \cite{2006A&A...457..841I,2008ApJ...684...88N,2012MNRAS.423..909C,2013MNRAS.431.1547B,2016MNRAS.460.4258L,2019ApJ...877..117H,2018MNRAS.478..592H,2019MNRAS.489..820B,2020MNRAS.491.4768R,2019ApJ...881...80L,2019MNRAS.483.2487J,2019MNRAS.483.2801S,2019arXiv191007127A,2020A&A...637A.100W,2004.09542}, but for the foreseeable future, reliable methods to marginalize over uncertainties in the $N(z)$ will be crucial to control residual systematic errors and fully propagate this uncertainty to final parameter constraints.

    The traditional approach has been to add nuisance parameters that shift the mean of the best guess $N(z)$ or, additionally, change its width or a relative contribution from a secondary peak in $N(z)$ (e.g. \cite{2018PhRvD..98d3526A,0912.0201}). This approach has so far been sufficient but, in addition to adding a potentially large number of nuisance parameters, the method suffers from a model completeness problem: how do we know that any particular parametrization encompasses all possible ways in which our best-guess $N(z)$ can be wrong?

    In this paper we propose a new technique that performs a marginalization around all possible functional deviations from the best-guess $N(z)$. The space of possible deviations in $N(z)$ is described as a Gaussian function and, under some controlled approximations, these deviations can be marginalized over analytically. This results in a simple change to the data covariance matrix with a simple explanation: directions in the space of predictions that are degenerate with changes in $N(z)$ are given large variance, which means that information in these directions is not used for inferring the parameters of interest. The method then marginalizes over the $N(z)$ uncertainty in a robust manner, without including a large number of new nuisance parameters. {A similar approach for analytically marginalizing over calibration and beam uncertainties in order to obtain fast and accurate constraints on cosmological parameters has been proposed for performing cosmic microwave background (CMB) data analysis (see \cite{2002MNRAS.335.1193B}).}

    The paper is structured as follows. In Section \ref{sec:theory} we derive the basic equations and discuss how the priors on the allowed fluctuations should be determined. In Section \ref{sec:hsc} we  apply this method to the HSC first public data release as an example to demonstrate its applicability in a practical settings. We conclude in Section \ref{sec:conclusions} and discuss some of the advantages, limitations and possible extensions of this model.

  \section{Theory}\label{sec:theory}
    \subsection{Redshift distribution uncertainties}\label{ssec:theory.nz}
      Most cosmological analyses involve performing Bayesian parameter inference on a posterior distribution $p(\vec{\theta} |\vd) \propto p(\vd|\vec{\theta}) p(\vec{\theta})$, where $\vd$ is a vector of data points and $\vec{\theta}$ is a set of parameters describing the underlying model.
      
      In the case of tomographic large-scale structure analyses, $\vd$ traditionally contains correlation function or power spectrum measurements between different tracers (e.g. galaxy ellipticities or overdensity) and different redshift bins\footnote{We use the term ``redshift bin'' here to denote any given galaxy sample, whether or not it has actually been selected by binning those galaxies into intervals of some redshift estimate.}, and $\vec{\theta}$ includes both cosmological and nuisance parameters needed to produce a forward model of $\vd$ that we will call $\vt(\vec{\theta})$. Due to the central limit theorem it is often accurate enough to assume that the likelihood is Gaussian {\cite{2018MNRAS.473.2355S,2019arXiv190503779L,2020arXiv200510384B}}, taking the form
      \begin{equation}
        \mathcal{L} \equiv p({\bf d}|\vec{\theta}) = \frac{\exp\left [-\frac{1}{2}(\vd-\vt(\vec{\theta}))^T\sC^{-1}(\vd-\vt(\vec{\theta})) \right]}{\sqrt{\det{2\pi\sC}}}, \label{eq:like}
     \end{equation}
      where $\sC$ is the covariance matrix of $\vd$. The information contained in the parameter dependence of $\sC$ can be neglected \citep{1811.11584}, and therefore the normalization factor $\sqrt{2\pi\det\sC}$ can be ignored.

      We can make significant progress without specifying explicitly how the theory is calculated given the parameters. It suffices to say that, given a set of cosmological parameters, which specify the expansion history of the universe and the growth of matter inhomogeneities as a function of scale and redshift, a set of astrophysical parameters (e.g. a particular bias or intrinsic alignment model), and the redshift distributions $N_i(z)$ of the different redshift bins, we can integrate these into theory predictions $\vt=\vt(\vec{\theta})$. Our main concern in this paper is the way in which the significant uncertainties in $N(z)$, which can potentially dominate the total systematic error budget, should be parametrized and marginalized over.

      The traditional approach (e.g. \cite{2018PhRvD..98d3526A,2019PASJ...71...43H,1912.08209,2020A&A...637A.100W}) has been to parametrize departures from a given ``educated guess'' of the redshift distribution that are likely to describe the main modes in which the underlying uncertainty propagates into the theory vector. This has usually been done by introducing ``shift'' ($\Delta z$) and/or ``width'' ($z_w$) nuisance parameters, in terms of which the fiducial redshift distribution for the $i$-th redshift bin $\bar{N}_i(z)$ is modified as
      \begin{equation}
        N_{i}(z) \propto \bar{N}_{i}\big(z_{c,i} + (1 + z_{w, i})(z-z_{c, i}) + \Delta z_{i}\big), \label{eq:photo-z-model}
      \end{equation}
      where $z_{c,i}$ is the mean of the fiducial distribution and acts as a sensible pivot to define width variations. $\Delta z_i$ and $z_{w,i}$ for each bin are then inferred and marginalized over together with all other parameters, with priors on them based on calibration uncertainties.

      This approach explicitly marginalizes over two effects that are deemed to be the most important systematics resulting from the redshift distribution uncertainties: the width affects the amplitude and shape of the clustering as well as the shear cross-power spectra, while the shift varies the distance to the sample in the case of clustering, and overlap with the lensing kernel in the case of shear cross-correlations. For shear power spectra, the effects are less pronounced given the broad size of the lensing kernel but nevertheless important for future experiments.  However, this methods suffers from two shortcomings. First, it is fundamentally ad-hoc: while this particular parametric form intuitively characterizes the most important effects, it is not directly mapped onto the main modes of uncertainty in photometric redshift estimators, and there is no rigorous proof that the parametrization is complete in the sense that it captures all relevant effects. The second problem is that it is computationally expensive. By adding one or two additional nuisance parameters per redshift bin, the dimensionality of the model parameter space increases significantly. For instance, in the analysis of \cite{1912.08209}, redshift uncertainty parameters accounted for more than half of all model parameters. The methodology described in the next section aims to address both of these shortcomings.

    \subsection{A new approach}\label{ssec:theory.nz_new}
      Let us consider a discretized description of the $N_i(z)$ as a sum over basis functions $\phi_\alpha(z)$:
      \begin{equation}\label{eq:n_discrete}
        N_i(z)=\sum_i N^\alpha_i\phi_\alpha(z).
      \end{equation}
      For simplicity here we will just treat $N_i(z)$ as a histogram, and therefore in our case the $\phi_\alpha$ are just top-hat functions centered on an equi-spaced grid of redshifts $z_\alpha$. The formalism described below is however valid for any choice of basis functions. In particular, it is likely that future photometric estimators will provide an estimate of uncertainty expressed as variance in a few dominating modes (PCA-like components) and specify the corresponding covariance matrix. We describe the full set of redshift distributions by the vector $\vN\equiv\{N^\alpha_i\,\,\,\forall i,\alpha\}$.  The normalizations of the $N_{i}(z)$ are irrelevant, since they only enter the theory prediction for the power spectrum as normalized redshift probability distributions (see Eq. \ref{eq:cell_gg_limber}).

      Let us further distinguish between the true underlying redshift distribution, represented by $\vN$, and a possible measurement of it $\hat{\bf N}$, with measurement uncertainties encapsulated by a covariance matrix $\hat\sC_N$. Furthermore, we may have some external prior information on $\vN$, which for simplicity we will assume to be Gaussian with mean $\vN_P$ and covariance $\hat\sC_P$.
      
      Our data is therefore made up of the combination ${\bf d}=(\hat{\bf c},\hat{\bf N})$, where $\hat{\bf c}$ is a set of two-point function measurements. The model parameters are $\vec{\theta}=({\bf q},\vN)$, where $\vN$ are the redshift distribution coefficients, and ${\bf q}$ contains all other cosmological, astrophysical and nuisance parameters. The posterior distribution is therefore given by:
      \begin{align}
        p(\vec{\theta}|{\bf d})
        &=p({\bf q},\vN|\hat{\bf c},\hat{\bf N})\\\label{eq:posterior_full}
        &\propto p(\hat{\bf c}|{\bf q},\vN)p(\hat{\bf N}|\vN)p(\vN)p({\bf q})\\
        &=p({\bf q})\frac{\exp\left[-\frac{1}{2}(\chi^2_c+\chi^2_N+\chi^2_P)\right]}{\sqrt{{\rm det}(2\pi\sC_c){\rm det}(2\pi\sC_N){\rm det}(2\pi\sC_P)}},
      \end{align}
      where
      \begin{align}
        \chi^2_c&=\left(\hat{\bf c}-{\bf t}\right)^T\sC_c^{-1}\left(\hat{\bf c}-{\bf t}\right)\\
        \chi^2_N&=\left(\hat{\bf N}-\vN\right)^T\sC_N^{-1}\left(\hat{\bf N}-\vN\right)\\
        \chi^2_P&=\left(\vN-\vN_P\right)^T\sC_P^{-1}\left(\vN-\vN_P\right),
      \end{align}
      and ${\bf t}({\bf q},\vN)$ is the theoretical prediction for $\hat{\bf c}$.
      
      In Eq. \ref{eq:posterior_full}, we have assumed that $\hat{\bf c}$ and $\hat{\vN}$ are independent at the likelihood level, and that the likelihood of $\hat{\vN}$ is independent of ${\bf q}$. The validity of these assumptions should be studied in detail, especially in cases where $\vN$ is calibrated using a spectroscopic sample with significant spatial overlap with the data under study (through a clustering redshift approach \cite{2008ApJ...684...88N} or otherwise). The method outlined here is still applicable if these assumptions are dropped, albeit with a different expression for the modified covariance in Eq. \ref{eq:main2}. We have also assumed that the deviations from the best guess redshift distribution $\hat{\vN}$ are Gaussianly distributed \cite{2004.09542}. While this is not true in detail for current redshift distributions, $N(z)$, it will likely become a better approximation in the future, as more spectroscopic data becomes available. Moreover, this is also unlikely to  matter much in practice, as long as $\sC_N$ captures the relevant directions and amplitudes of uncertainty.   

      To proceed further, let us start by considering the combination $\chi^2_N+\chi^2_P$. After completing squares for $\vN$, this can be written as:
      \begin{equation}
        \chi^2_N+\chi^2_P\equiv\chi^2_{\bar{N}}=\left(\vN-\bar{\vN}\right)^T{\sf P}^{-1}\left(\vN-\bar{\vN}\right)+K_N,
      \end{equation}
      where we have defined the \emph{smoothed mean} $\bar{\vN}$ and combined prior covariance ${\sf P}$ as:
      \begin{align}\label{eq:priorcov}
        \bar{\vN}\equiv{\sf P}(\sC_N^{-1}\hat{\bf N}+\sC_P^{-1}\vN_P),\hspace{12pt} {\sf P}^{-1}\equiv\sC_N^{-1}+\sC_P^{-1},
      \end{align}
      and $K_N$ is independent of the model parameters, and given by
      \begin{equation}
        K_N\equiv\hat{\bf N}^T\sC^{-1}_N\hat{\bf N}+\vN_P^T\sC_P^{-1}\vN_P-\bar{\vN}^T{\sf P}^{-1}\bar{\vN}.
      \end{equation}
      The full posterior thus takes the form:
      \begin{align}
        p({\bf q},\vN|\hat{\bf c},\hat{\bf N})&=Q\,p({\bf q})\exp\left[-\frac{1}{2}(\hat{\bf c}-{\bf t})^T\sC^{-1}_c(\hat{\bf c}-{\bf t})-\frac{1}{2}(\vN-\bar{\vN})^T{\sf P}^{-1}(\vN-\bar{\vN})\right],\\
        Q&\equiv\frac{e^{-K_N/2}}{\sqrt{{\rm det}(2\pi\sC_c){\rm det}(2\pi\sC_N){\rm det}(2\pi\sC_P)}}.
      \end{align}

      We can now write a likelihood for $\hat{\bf c}$ in which we explicitly marginalize out all the degrees of freedom associated with $\vN$:
      \begin{equation}
        \mathcal{L} \propto  \int \exp\left [-\frac{1}{2} (\hat{\bf c}-\vt)^T {\sf C}_c^{-1} (\hat{\bf c}-\vt) -\frac{1}{2} (\vN - \bar{\vN})^T {\sf P}^{-1} (\vN -\bar{\vN}) \right]  d^N\vN, \label{eq:like2}
      \end{equation}
      Note that in the equation above, $\vt=\vt({\bf q},\vN)$. It is the non-linear dependence of $\vt$ on $\vN$ that prevents us from performing what looks like a trivial integral analytically. In principle, we could solve this problem by making every single element of the vector $\vN$ a part of an MCMC chain for ${\bf q}$ and additional hundreds of $\vN$ parameters. This should be feasible using methods that can sample very large parameter spaces, such as Hamiltonian Monte Carlo \cite{1987PhLB..195..216D}. Instead, faced with queue waiting times at NERSC, we Taylor expand the theory in $\vN$ to first order around $\bar{\vN}$:
      \begin{equation}
        \vt({\bf q},\vN) \simeq \vt({\bf q},\bar{\vN}) + {\sf T} \left(\vN - \bar{\vN} \right), \label{eq:taylor}
      \end{equation}
      where the matrix ${\sf T}$ is the gradient of ${\bf t}$ with respect to $\vN$
      \begin{equation}
        {\sf T}\equiv \left.\frac{d{\bf t}}{d\vN}\right|_{{\bf q},\bar{\vN}}.
        \label{eq:Tmat}
      \end{equation}
      In other words, ${\sf T}$ contains the response of all the correlation functions or power spectra to small changes in $N_i(z)$.

      Substituting Eq. \ref{eq:taylor} into Eq. \ref{eq:like2} results in an integral that is now quadratic in $\vN$ and can be performed analytically. After completing squares, and applying the Woodbury matrix identity\footnote{\url{https://en.wikipedia.org/wiki/Woodbury_matrix_identity}}, we find:
      \begin{equation}
        \mathcal{L} \propto \left[{\rm det}\left({\sf T}^T \sC_c^{-1} {\sf T} +{\sf P}^{-1}\right)\right]^{-1/2} \exp\left [-\frac{1}{2} ({\bf c}-\vt)^T \sC_M^{-1} ({\bf c}-\vt) \right], \label{eq:main1}
      \end{equation}
      where the  marginalized covariance is
      \begin{equation}
        \sC_M = \sC_c + {\sf T}{\sf P}{\sf T}^T. \label{eq:main2}
      \end{equation}
      This is a fascinatingly simple equation and the main result of this paper. This calculation can be understood as follows: for each deviation around $\bar{\vN}$, there is a corresponding deviation around $\vt$. These are directions in the space of theory predictions that are perfectly degenerate with the changes in shape of $N_i(z)$.  Equation \ref{eq:main2}  increases the variance for those linear combinations commensurately with how far ${\sf P}$ allows them to go. The limit ${\sf P}\rightarrow\infty\mathbb{1}$ would correspond to completely projecting out those ``$N(z)$-sensitive'' modes from the data altogether. 

      Both the marginalized covariance $\sC_M$, and the normalizing prefactor in Eq. \ref{eq:main1} depend in principle on ${\bf q}$ through the parameter dependence of ${\sf T}$. This implies that, in principle, ${\sf T}$ should be re-evaluated at every point in the MCMC chain,  when sampling the likelihood in Eq.~\ref{eq:main1}. Since the calculation of ${\sf T}$ is expensive using standard methods (by comparison with that of e.g. $\vt$), we will neglect this dependence and evaluate ${\sf T}$ at a fiducial set of parameters. We will however explore the impact of the choice of fiducial parameters on the final results. This should be a sufficiently good approximation for compact likelihoods, where parameter constraints are driven by the differences between $\hat{\bf c}$ and $\vt$. Computing ${\sf T}$ for a fiducial set of parameters implies that, for a fixed data covariance matrix $\sC_c$, the normalization prefactor in Eq.~\ref{eq:main1} is an irrelevant constant, and the marginalized covariance in Eq. \ref{eq:main2} also needs to be evaluated only once.

      The modified covariance matrix in effect artificially increases the variance for certain linear combinations of the data. Therefore, the effective number of degrees of freedom will be lower. We can estimate this effect by noting that
      \begin{equation}
        \mbox{d. o. f.} \equiv \left<({\bf c}-{\bf t})^T \sC_M^{-1} ({\bf c}-{\bf t}) \right> = {\rm Tr}\left( \sC_c \sC_M^{-1} \right),
        \label{eq:dof}
      \end{equation}
      where we have assumed the data to be distributed according to the original, unmodified covariance. If the data are however actually contaminated according to the model for $N(z)$ uncertainties described here, the actual expected degrees of freedom will be given by the original number of data points. Therefore, depending on how aggressive we are with the marginalization, the actual degrees of freedom used in  $\chi^2$ test will lie between the result of Equation \ref{eq:dof} and true number of data points.
      
      In order to obtain constraints from the marginalized likelihood in Eq.~\ref{eq:main1}, we can proceed as follows:
      \begin{itemize}
        \item Solve the parameter inference problem once with fixed redshift distribution and a standard Gaussian likelihood for $\hat{\bf c}$ with covariance $\sC_c$ to determine a sensible fiducial model.
        \item Calculate ${\sf T}$ and $\sC_M$ at this fiducial model.
        \item Solve again the inference problem with a standard Gaussian likelihood using the modified covariance matrix $\sC_M$ instead of $\sC_c$. This will lead to broadened contours due to the marginalization over redshift distribution uncertainties.
      \end{itemize}
      If needed, one could repeat the inference at a refined fiducial model although in practice we found this to be unnecessary.

      Before moving on, it is worth emphasizing the two distinct approximation used here. The first one is that the theory can validly be Taylor expanded in $\vN-\bar{\vN}$. The second is that the model dependence of ${\sf T}$ can be ignored for the models of interest. We will examine the validity of both approximations in Section \ref{sec:hsc}.
      
    \subsection{The prior matrix ${\sf P}$}\label{ssec:theory.prior}
      A key part of this method is the determination of the $N(z)$ prior covariance ${\sf P}$, which governs the amplitude of the uncertainties on $\vN$. As discussed in the previous section, ${\sf P}$ receives two contributions: the covariance associated with the uncertainties in the measured $\hat{\bf N}$, $\sC_N$, and the external prior with covariance $\sC_P$, both combined in an inverse-variance way (see Eq. \ref{eq:priorcov}). We will describe the models used for both contributions in this section.

      \subsubsection{$N(z)$ uncertainties}\label{sssec:theory.prior.cv}
        \begin{figure}[ht]
          \centering  
          \includegraphics[width=1.\textwidth]{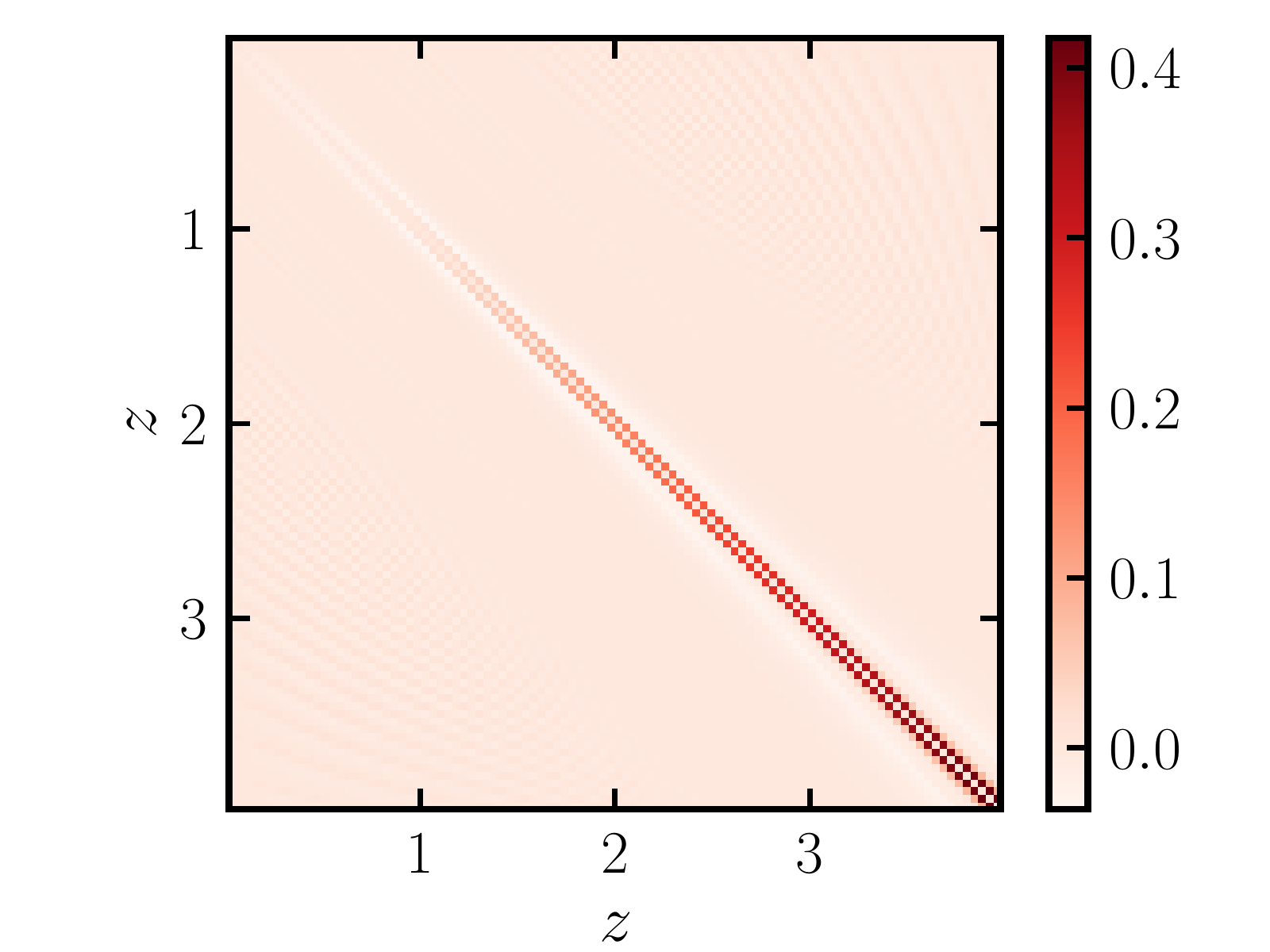}
          \caption{Correlation matrix associated with the cosmic-variance contribution to the total $N(z)$ covariance for the first tomographic bin of the HSC data.}\label{fig:CV}
        \end{figure}
        The analysis presented in Section \ref{sec:hsc} will make use of a measured redshift distribution estimated from the COSMOS 30-band catalog \cite{2016ApJS..224...24L}, as was done in \cite{1912.08209}. The main sources of uncertainty for this measurement are cosmic variance in the COSMOS field, shot noise and additional uncertainty in the photometric redshifts used to assign galaxies to different bins. Here we will associate each of these sources to a separate covariance that will be added in quadrature to form the final $\sC_N$. There is a fourth source of uncertainty that we do not address here: although the redshifts provided in the COSMOS 30-band catalog have a high accuracy, they are photometric and therefore subject to potential biases. This has been identified as an important source for differences between ongoing cosmic shear surveys \citep{2020A&A...638L...1J}. Calibrating this bias or incorporating it into the $N(z)$ error model is an important task that requires a detailed study of the COSMOS 30-band sample, and that we leave for future work.
        
        The sample variance uncertainties are caused by fluctuations in the matter density traced by galaxies in the particular sky patch. The corresponding covariance would ideally be estimated from simulations including both non-linear gravitational clustering and a realistic model of the galaxy-halo connection for the specific galaxy sample under study. For simplicity, in this proof-of-concept analysis, we instead use an analytical model for the $N(z)$ covariance. Although less precise than a simulation-based calculation, this approach was shown by \cite{2004.09542} to provide a reasonable prediction for redshift distribution uncertainties.

        The covariance matrix element between two tomographic bins $N^\alpha_i$ and $N^\beta_j$ is given by:
        \begin{equation}\label{eq:nz_cv}
          \sC_{N,{\rm CV}}^{(i\,\alpha),(j\,\beta)}=\frac{N^\alpha_iN^\beta_j}{2\pi^2} \int_0^\infty dk_\parallel \cos(k_\parallel(\chi_\alpha-\chi_\beta))\int_0^\infty dk_\perp k_\perp W_\alpha(k_\parallel,k_\perp)W_\beta(k_\parallel,k_\perp)P_{gg}({\bf k}),
        \end{equation}
        where $\chi_\alpha$ is the radial comoving distance to redshift $z_\alpha$, and we model the galaxy power spectrum using the Kaiser formula \cite{1987MNRAS.227....1K}, accounting for redshift-space distortions:
        \begin{equation}
          P_{gg}({\bf k},z)=\left(b_g(z)+f(z)\frac{k_\parallel^2}{k^2}\right)^2\,P_{mm}(k,z).
        \end{equation}
        Here, $f(z)$ is the logarithmic growth rate and $b_g(z)$ is the linear galaxy bias. Following the results of \cite{1912.08209}, appropriate for the magnitude-limited sample studied here, we assume a redshift dependence for $b_g$ given by $b_g(z) = 0.95/D(z)$, with $D(z)$ the linear growth factor. $P_{mm}(k)$ is the non-linear matter power spectrum, which we model using the revised HALOFIT parametrization \cite{2003MNRAS.341.1311S,2012ApJ...761..152T}.

        The part of HSC Deep\footnote{The HSC survey is subdivided into three different parts: Wide, Deep and UltraDeep. The Deep part of the survey covers approximately 27 square degrees to a limiting $i$-band magnitude of $m_{\mathrm{lim}, i} \sim 27$ \cite{2018PASJ...70S...8A}.} data which overlaps with the COSMOS 30-band footprint and is used to measure the redshift distribution in this work, covers a total area of $A_{\rm sky} = 1.7 \ {\rm deg}^2$. Modeling this patch as a disc of radius $\theta_{\rm sky}=0.73^\circ$, the corresponding window function in Eq. \ref{eq:nz_cv} is given by
        \begin{equation}
          W_\alpha(k_\parallel,k_\perp)=j_0(k_\parallel\Delta\chi_\alpha/2)\,\frac{2 J_1(k_\perp \chi_\alpha\theta_{\rm sky})}{k_\perp \chi_\alpha\theta_{\rm sky}},
        \end{equation}
        where $\Delta\chi_\alpha$ is the comoving width of the $\alpha$-th redshift histogram bin, $j_0$ is the zero-th order spherical Bessel function and $J_1$ is the order-1 cylindrical Bessel function.

        We assume that the cosmic covariance between the tomographic bins is negligible, so we estimate the cosmic variance covariance matrix by treating each bin independently. Fig. \ref{fig:CV} shows the correlation matrix associated with this cosmic variance contribution in the COSMOS 30-band sample for the first tomographic bin used in Section \ref{ssec:hsc.data}.
        
        The covariance matrix receives an additional contribution due to the Poisson noise, which is caused by the discrete nature of galaxies as a tracer of the matter fluctuations. This is given by
        \begin{equation}\label{eq:cov_nz_shot}
          \sC^{(i\,\alpha),(j\,\beta)}_{N,{\rm SN}}=\delta_{\alpha\beta}\delta_{ij}N^\alpha_i.
        \end{equation}
        We find this contribution to be subdominant in all cases in comparison with sample variance.

        Finally we estimate the systematic error associated with the choice of photo-$z$ code. Following \cite{1912.08209}, we consider  {\tt DEmP}, {\tt Ephor}, {\tt Ephor\_AB} and {\tt FRANKEN-Z} as some of the best-performing algorithms presented in \cite{2018PASJ...70S...9T}. We assume that the spread between these methods constitutes a fair representation of the underlying photo-$z$ uncertainties. Note however that all of these codes rely on COSMOS-30 for calibration. We calculate the variance between the $N(z)$s estimated from each code by stacking the redshift probability density functions (pdfs) of each source in the HSC data.  We multiply this variance by a factor $A_{\rm noise}$ and smooth it by convolving it with a Gaussian kernel with standard deviation $\sigma_z=0.1$. The resulting variance vector is added to the diagonal of the pure sample variance covariance matrix. We discuss the sensitivity of our analysis to some of these partially subjective choices in Section \ref{ssec:hcs.mcmc}.

    \subsubsection{External priors and smoothness}\label{sssec:theory.prior.smooth}
      In addition to obtaining information from direct measurements of the $N(z)$s, it is reasonable to impose certain properties of the underlying true redshift distributions, based on physical considerations. For instance, there is no reason to expect that the physics of galaxy formation should generate sharp features in the redshift evolution of the abundance of galaxies. The transition features in the spectra of different galaxy types between photometric bands at different redshifts could induce smaller-scale fluctuations in the $N(z)$. However, for a sufficiently diverse galaxy sample, one would not expect such fluctuations on scales $\delta z\lesssim0.04$ (corresponding to $\sim100\,{\rm Mpc}$ or $\sim0.2\,{\rm Gyr}$ at $z\sim1$), as these distances are comparable to redshift-space distortion smoothing. It is therefore a reasonable proposition to impose a certain degree of smoothness on the redshift distributions, which can be achieved through a purposely defined Gaussian prior. Applying these types of priors is admittedly a subjective choice to some extent, and we will study its impact on our results in Section \ref{ssec:hcs.mcmc}.
      
      One common way to impose smoothness on a function $f(x)$ is to penalize large values of its first derivative via a Gaussian prior of the form $p(f)\propto\sum_x\exp\left[-(f'(x))^2/(2\sigma_1^2)\right]$. In the discrete formalism used here, using first-order finite differences, this is equivalent to imposing a Gaussian prior on $\vN$ with zero mean ($\vN_P=0$) and an inverse covariance given by:
      \begin{equation}\label{eq:prior_1st}
        \sC^{-1}_P=\frac{1}{\sigma_1^2}\sum_\alpha {\bf v}^{1T}_\alpha{\bf v}^1_\alpha,
      \end{equation}
      where
      \begin{equation}
        ({\bf v}^1_\alpha)_\beta=\left\{
        \begin{array}{ll}
          1  & {\rm if}\,\,\, \alpha=\beta+1\\
          -1  &  {\rm if}\,\,\, \alpha=\beta\\
          0 & {\rm otherwise}
        \end{array}\right..
      \end{equation}
      To understand this, note that, using finite differences, the derivative of $\vN$ is approximately $\vN'\propto {\bf v}^{1T}\vN$.

      The prior in Eq. \ref{eq:prior_1st} effectively penalizes large deviations between adjacent elements of $\vN$. This can be generalized to include all possible pairs of elements as
      \begin{equation}\label{eq:smooth1}
        \sC^{-1}_P=\sum_{n=1}\sum_{\alpha}p_n\,{\bf v}^{n\,T}_\alpha{\bf v}^n_\alpha,
      \end{equation}
      where
      \begin{equation}
        ({\bf v}^n_\alpha)_\beta=\left\{
        \begin{array}{ll}
          1  & {\rm if}\,\,\, \alpha=\beta+n\\
          -1  &  {\rm if}\,\,\, \alpha=\beta\\
          0 & {\rm otherwise}
        \end{array} \right..
      \end{equation}
      The prefactor $p_n$ penalizes differences between neighbors of order $n$, and should therefore be a monotonically decreasing function of $n$ (since there should be no correlation between widely separated histogram bins). We therefore choose a functional form
      \begin{equation}
        p_n=A_{\rm smooth}\,{\rm exp}\left[-\frac{1}{2}\left(n\frac{\Delta z}{\Delta z_{\rm thr}}\right)^2\right],
      \end{equation}
      where $\Delta z$ is the redshift separation between neighboring histogram bins, and $\Delta z_{\rm thr}$ marks the redshift separation beyond which different elements of the $N(z)$ are expected to be uncorrelated. Our fiducial analysis uses $A_{\rm smooth}=1$ and $\Delta z_{\rm thr}=0.06$. These values were empirically chosen to cause a mild smoothing of the redshift distribution directly measured from the COSMOS catalog. We will return to choices of these parameters in Section \ref{ssec:hsc.lin}. As we discuss in Section \ref{ssec:hcs.mcmc}, this choice has no practical impact on the final results.
    
    \subsubsection{Up-sampling}\label{sssec:theory.prior.ups}
      Depending on the size of the spectroscopic sample used to estimate the initial redshift distribution, the measured $\hat{\vN}$ may be provided with a relatively coarse redshift spacing to reduce the jaggedness in the fiducial $N(z)$. However, it may often be desirable to explore the impact of variations in the $N(z)$ on scales smaller than that, which implies artificially increasing the size of $\vN$ by up-sampling the original distributions onto a finer grid of $z$.
      
      This up-sampling can be done in different ways, but in general can be expressed as a linear operation of the form
      \begin{equation}
        \vN_{\rm fine}={\sf O}\,\vN_{\rm coarse}.
      \end{equation}
      The prior covariance of $\vN_{\rm fine}$ is then related to that of $\vN_{\rm coarse}$ via the bilinear operation
      \begin{equation}
        {\sf P}_{\rm fine}={\sf O}\,{\sf P}_{\rm coarse}\,{\sf O}^T.
      \end{equation}
      
      For nearest-neigh interpolation, the linear kernel $O_{\mu\alpha}$ is simply $O_{\mu\alpha}=\delta_{\alpha\alpha_\mu}$, where $\alpha_\mu$ is the index of the coarse redshift distribution element that lies closest to the finer grid element with index $\mu$. Higher-order interpolation methods can be described in terms of different kernels. In practice, the easiest procedure is to simply apply the same interpolating function used to up-sample $\vN$ to all the rows and then all the columns of ${\sf P}_{\rm coarse}$.
      
      The original $N(z)$s obtained from the COSMOS catalog were measured in bins of $\Delta z=0.04$ in the range $z\in(0,4)$. We up-sampled them using linear interpolation by a factor $N_{\rm up}=3$ to a resolution $\Delta z=0.0133$. For the 4 bins used in this analysis, the final up-sampled $\vN$ has 1200 elements. We note that it is important that the sample-variance calculation is performed on the original binning, since this is the binning over which the $N(z)$s were determined, and then up-sampled using the same linear operator.

  \section{Application to galaxy clustering in HSC}\label{sec:hsc}
    We apply the method outlined in the previous section to the data presented in Ref.~\cite{1912.08209}. This work measured the angular galaxy clustering power spectrum from the first data release (PDR1) of the Hyper Suprime-Cam survey, described in detail in \cite{2018PASJ...70S...8A,2018PASJ...70S..25M,2018PASJ...70S...5B}. In the following, we give a very brief summary of the methodology employed in Ref.~\cite{1912.08209} as well as the dataset being used and refer the reader to the original paper for further details. Next, we explore the accuracy of the approximations adopted in the $N(z)$ marginalization prescription introduced in the previous section. Finally, we study the impact of our method on the inferred cosmological constraints, comparing it with the analysis in Ref.~\cite{1912.08209} and doing some consistency checks to test its robustness.
    
    \subsection{Background theory}\label{ssec:hsc.theory}
      The angular clustering power spectrum for galaxies in redshift bins $i$, $j$ can be modeled using the Limber approximation as \cite{1953ApJ...117..134L, 1992ApJ...388..272K, Kaiser:1998}
      \begin{equation}\label{eq:cell_gg_limber}
        C^{ij}_\ell = \int \mathrm{d}z\,\frac{H(z)}{\chi^2(z)} p^i(z)p^j(z)\,P_{gg}\left(z,k=\frac{\ell+1/2}{\chi(z)}\right),
      \end{equation}
      where $P_{gg}(z,k)$ denotes the underlying 3D galaxy power spectrum, $\chi(z)$ is the comoving distance and $H(z)$ denotes the Hubble parameter at redshift $z$. $p^i(z)$ is the redshift probability distribution of bin $i$ normalized to unit area, and is therefore related to the unnormalized distribution via
      \begin{equation}
        p^i(z)=\frac{N_i(z)}{\int dz' N_i(z')}=\frac{\sum_\alpha N^\alpha_i\phi_\alpha(z')}{\sum_\alpha N^\alpha_i\int dz\phi_\alpha(z')}.
      \end{equation}
      The simplicity with which the redshift distribution amplitudes $N_i^\alpha$ enter the prediction for the angular power spectrum in Eq.~\ref{eq:cell_gg_limber} {as linear and quadratic factors (also exemplified in Appendix \ref{app:autos})}, facilitates the computation of the ${\sf T}$ matrix defined in Eq. \ref{eq:Tmat}. 

      Following Ref.~\cite{1912.08209} we estimate the theoretical prediction for the galaxy power spectrum $P_{gg}(z,k)$ within the halo model combined with halo occupation distribution (HOD) modeling \cite{2000MNRAS.318.1144P,2002PhR...372....1C,2002ApJ...575..587B,2005ApJ...633..791Z,2013MNRAS.430..725V}. Details about HOD parametrizations can be found in these references, and here we only provide a succinct description relevant to the present analysis.

      The galaxy power spectrum receives contributions from the so-called 1-halo and 2-halo terms:
      \begin{equation}
        P_{gg}(z,k) = P_{gg,{\rm 1h}}(z,k) + P_{gg,{\rm 2h}}(z,k),
      \end{equation}
      where
      \begin{align}
        & P_{gg,{\rm 1h}}(k)=\frac{1}{\bar{n}_g^2} \int \mathrm{d}M\,\frac{\mathrm{d}n}{\mathrm{d}M} \bar{N}_c\,\left[\bar{N}_s^2u_s^2(k)+2\bar{N}_su_s(k)\right],\\
        & P_{gg,{\rm 2h}}(k)=\left(\frac{1}{\bar{n}_g} \int \mathrm{d}M\,\frac{\mathrm{d}n}{\mathrm{d}M}\,b_h(M)\,\bar{N}_c\,\left[1+\bar{N}_su_s(k)\right]\right)^2\,P_{\rm lin}(k).
      \end{align}
    
      Here, $\mathrm{d}n/\mathrm{d}M$ is the halo mass function for halo mass $M$, $b_h(M)$ is the linear halo bias, $\bar{N}_c(M)$ and $\bar{N}_s(M)$ are the mean number of central and satellite galaxies in halos of mass $M$, $u_s(k)$ is the Fourier transform of the satellite density profile, and $P_{\rm lin}(k)$ is the linear matter power spectrum. The number density of galaxies is calculated as 
      \begin{equation}
        \bar{n}_g=\int \mathrm{d}M\,\frac{\mathrm{d}n}{\mathrm{d}M}\bar{N}_c(M)\left[1+\bar{N}_s(M)\right].
        \label{eq:ng_hod}
      \end{equation}    
    
      As in \cite{1912.08209}, we parametrize the number of centrals and satellites as a function of mass as:
      \begin{align}
        &\bar{N}_c(M)=\frac{1}{2}\left[1+{\rm erf}\left(\frac{\log_{10}(M/M_{\rm min})}{\sigma_{\log M}}\right)\right],\\
        &\bar{N}_s(M)=\Theta(M-M_0)\left(\frac{M-M_0}{M_1'}\right)^\alpha,
      \end{align}
      where $\Theta(x)$ is the Heavyside step function, and we model the distribution of satellites to follow that of the dark matter, given by a truncated Navarro-Frenk-White profile \cite{Navarro:1996}. The choices of mass function parametrization, halo bias and concentration-mass relation used here  follow the same models used in \cite{1912.08209}, namely the mass function and halo bias of \cite{Tinker:2010}, and the concentration-mass relation of \cite{Duffy:2008} for spherical overdensity halo masses with an overdensity parameter $\Delta=200$ with respect to the critical density.
 
      The HOD model is defined by three characteristic masses, $M_{\rm min}$, $M_0$ and $M_1$. We model the redshift dependence of these masses as a linear Taylor expansion in the scale factor around the mean redshift of the sample $z_p=0.65$ as
      \begin{equation}
        \log_{10}{M_x(z)} = \mu_x + \mu_{x, p} \left(\frac{1}{1+z} - \frac{1}{1+z_{p}}\right),
      \end{equation}
      where $x$ is $\mathrm{min}$, 0 or 1.
    
      Besides these HOD parameters, the analysis of \cite{1912.08209} marginalized over uncertainties in the redshift distribution using the shift-width parametrization of Eq. \ref{eq:photo-z-model}, which results in including two additional shift and width parameters per redshift bin. For the four redshift bins used in this analysis, the complete set of 14 free parameters is thus
      \begin{equation}
        \vec{\theta}=\{\mu_{\rm min},\,\mu_{{\rm min},p},\,\mu_0,\,\mu_{0,p},\,\mu_1,\,\mu_{1,p},\,\Delta z_{\{1,2,3,4\}},\,z_{w,\{1,2,3,4\}}\}.
      \end{equation}

    \subsection{The HSC dataset}\label{ssec:hsc.data}
      The Hyper-Suprime Cam survey is an on-going photometric galaxy survey survey focused mainly on weak gravitational lensing. The analysis in Ref.~\cite{1912.08209} is based on the publicly-available HSC DR1 data \cite{2018PASJ...70S...8A}, whose so-called wide fields cover approximately 108 square degrees on the sky, subdivided into seven distinct patches. Ref.~\cite{1912.08209} used these data to compute spherical harmonic galaxy clustering power spectra for four tomographic redshift bins between $z=0.15$ and $z=1.5$, taking both auto- and cross-correlations into account. These power spectra have been corrected for observational and extragalactic systematics by deprojection at the map-level \cite{2019MNRAS.484.4127A}, and we have applied the power spectrum scale cuts described in \cite{1912.08209}. Finally, following Ref.~\cite{2019PASJ...71...43H}, photometric redshift distributions have been estimated by cross-matching HSC galaxies to galaxies in the COSMOS 30-band photometric catalog presented in Ref.~\cite{2016ApJS..224...24L}.

    \subsection{Validating the linear expansion}\label{ssec:hsc.lin}
      \begin{figure}[ht]
        \centering
        \includegraphics[width=1.\textwidth]{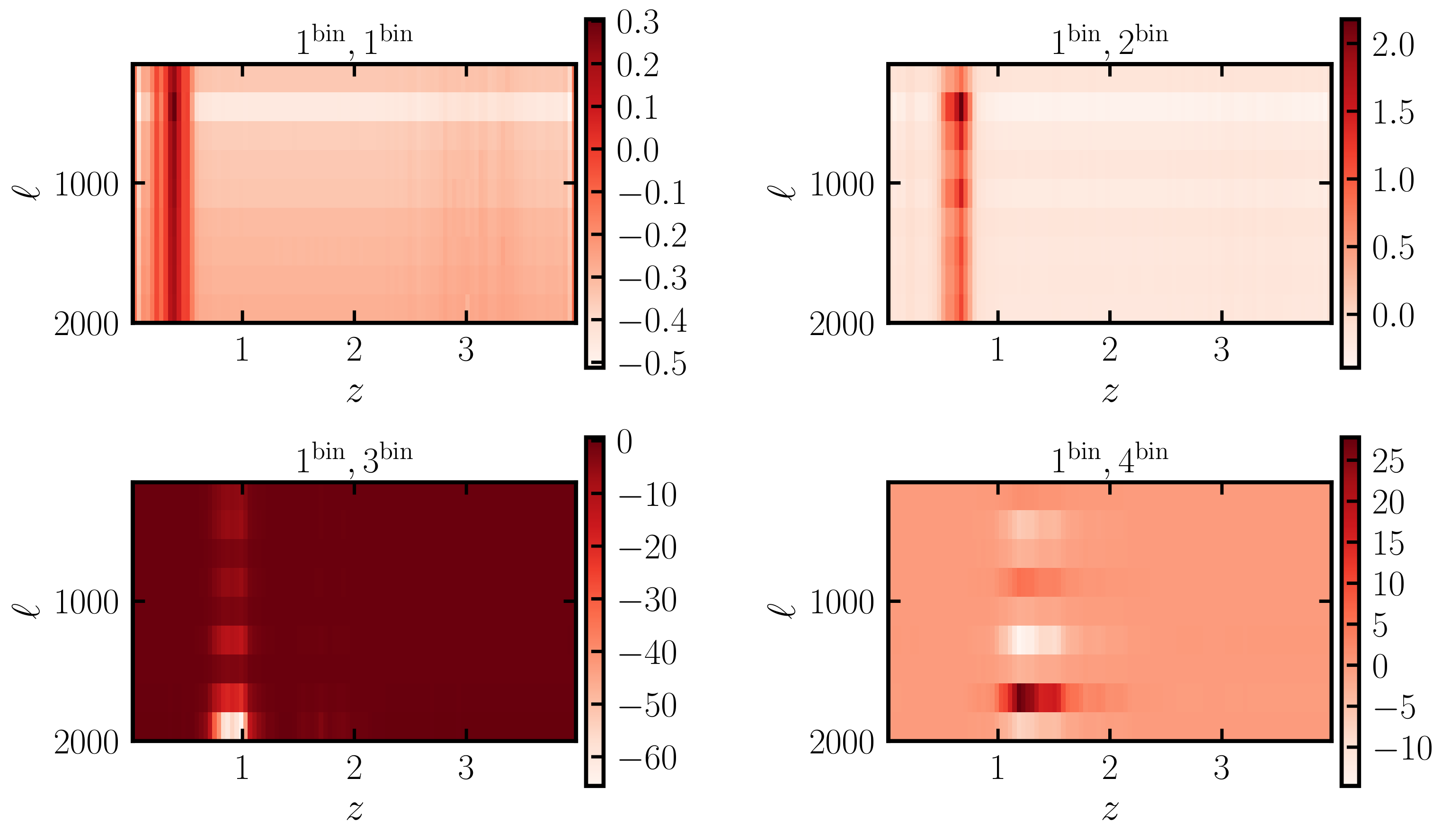}
        \caption{Visualization of the linear derivative matrix ${\sf T} = {\partial \vt}/{\partial \vN}$ divided by the angular power spectrum $C_{\ell}$ for all cross-correlation pairs involving the first tomographic bin.} \label{fig:Tmat}
      \end{figure}

      \begin{figure}[ht]
        \centering
        \includegraphics[width=1.\textwidth]{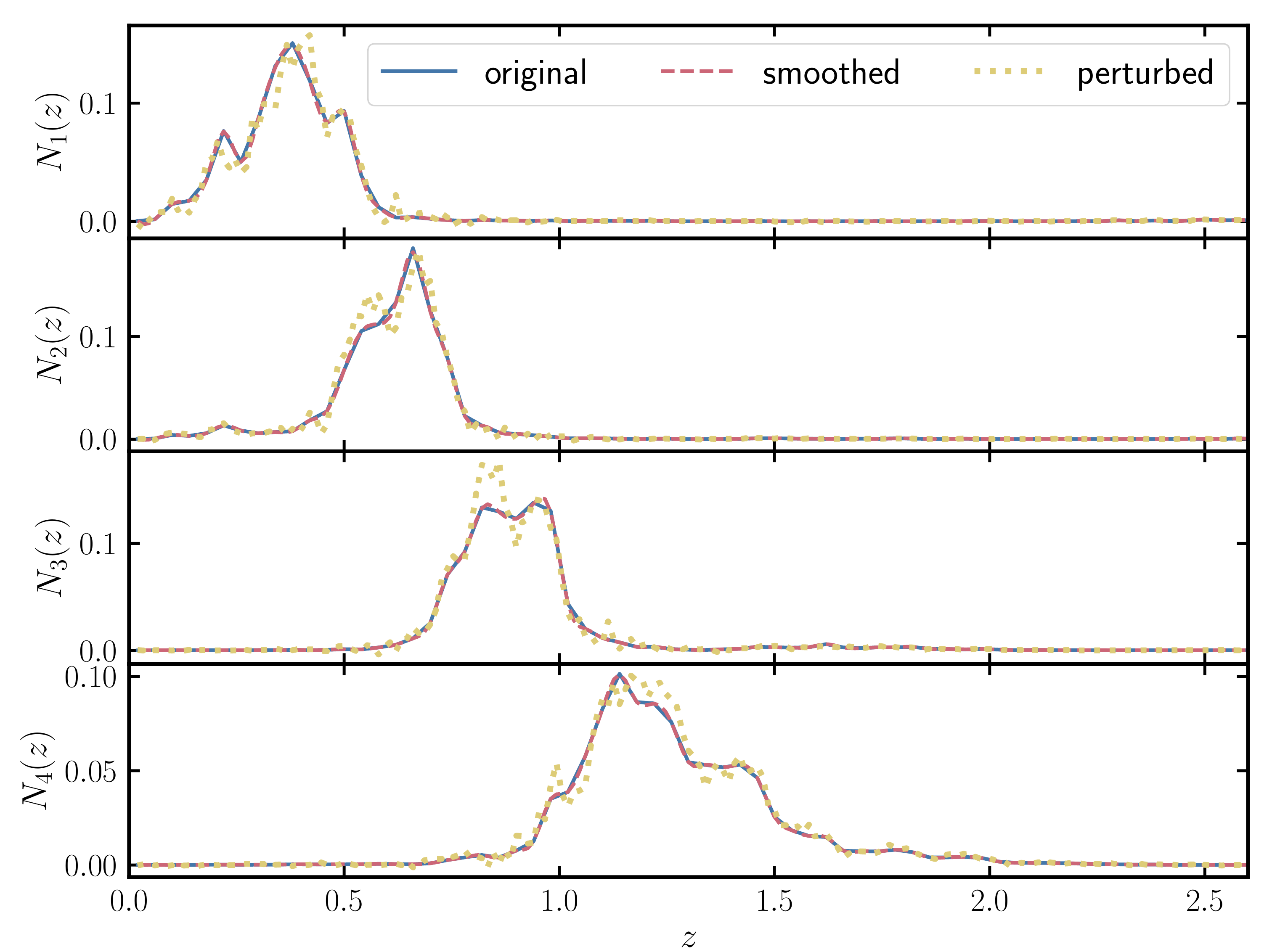}
        \caption{Photometric redshift distributions $N(z)$ of the four tomographic HSC bins. The \textit{blue solid} curves show the default values used in the full HSC analysis \cite{1912.08209}, while the \textit{red dashed} curves show their smoothed and up-sampled version used in this work. The \textit{thick yellow dotted} curves correspond to a random Gaussian draw from the smoothing prior described in Section~\ref{sssec:theory.prior.smooth}.}\label{fig:Nzs}
      \end{figure}

      \begin{figure}[ht]
        \centering
        \includegraphics[width=1.\textwidth]{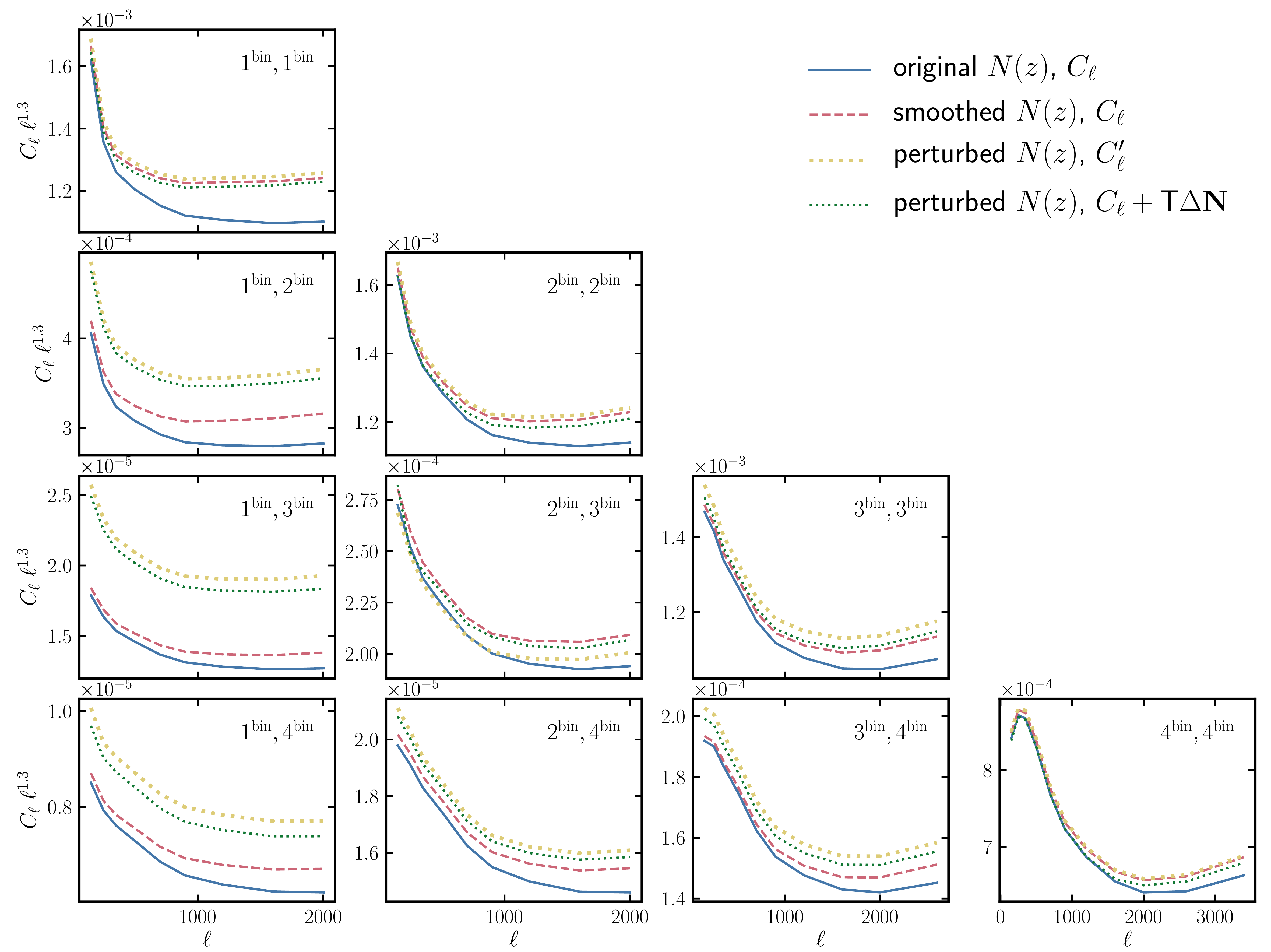}
        \caption{Galaxy power spectrum with subtracted shot noise, $C_\ell^{gg}$, between the four tomographic bins. The \textit{blue} {\textit{solid}} curves show the $C_\ell$s computed using the posterior mean values of the 6 HOD parameters from the full HSC analysis \cite{1912.08209}, while the \textit{red} {\textit{dashed}} curves show the angular power spectrum obtained by using the smoothed $N(z)$ distributions (see Fig.~\ref{fig:Nzs}). The \textit{thick yellow dashed} curves correspond to the exact calculation of the $C_\ell$s using the perturbed $N(z)$ values in Fig.~\ref{fig:Nzs}, and the \textit{light-blue} dotted lines show the power spectra derived via Taylor expansion: $C_\ell' = C_\ell + {\sf T} \left(\vN' - {\vN} \right)$. {Visually, the two curves exhibit a very good level of agreement. This finding is further corroborated by Table~\ref{tab:lin}, where we demonstrate that the $\chi^2$ values in both cases are remarkably similar}.}\label{fig:Cls}
      \end{figure}

      In this section, we study how the linear approximation performed via the derivative matrix ${\sf T}$ (defined in Eq. \ref{eq:Tmat}) compares with the exact $C_\ell$ calculation. In particular, we compare the original photometric redshift distributions as well as the angular power spectra with those obtained using the smoothing technique developed in Eq. \ref{eq:priorcov}, commenting on their ability to provide fits to the HSC data.
      
        We start by fixing the values of all the shift and width parameter to be zero and setting the 6 HOD parameters to their fiducial values of
      \begin{equation}\label{eq:par_fid}
        (\mu_{\rm min},\mu_{{\rm min},p},\mu_0,\mu_{0,p},\mu_1,\mu_{1,p})=(11.88,-0.5,5.7,2.5,13.08,0.9).
      \end{equation}
      These HOD values correspond to the posterior means reported in Ref. \cite{1912.08209}.
      
      The derivative matrix can be seen in Fig.~\ref{fig:Tmat} for the first tomographic bin and its cross-correlations with all four bins in the HSC data, i.e. the tomographic pairs [1,1], [1,2], [1,3], and [1,4]. The rows of the matrix correspond to the $\ell$ multipole bandpowers available for each pair, and the columns to the 300 equally-spaced redshift samples of $N(z)$ between $z = 0$ and $z = 4$ for each tomographic bin. We notice that as we move to pairs with bins at higher redshifts, the corresponding block of ${\sf T}$ peaks at a redshift column roughly corresponding to the maximum of the deepest distribution of the pair.
      
      The corresponding $N(z)$s are shown in Fig.~\ref{fig:Nzs}. As outlined above and described in Ref.~\cite{1912.08209}, we obtain the fiducial ``measured'' distributions $\hat{\vN}$, shown in solid blue in the figure, using the COSMOS 30-band catalog re-weighted to account for the different color space distribution of that sample using a nearest-neighbour approach. The figure also shows the {\sl smoothed} distribution $\bar{\vN}$ in dashed red, defined in Eq. \ref{eq:priorcov}. Our choice of smoothing prior has only a mild effect on the original redshift distribution, mostly removing sharp features at the edges of adjacent $N(z)$ measurements. Nevertheless, these small changes in $N(z)$ are sufficient to affect the effective large-scale bias of the theory prediction {(as demonstrated in Fig.~\ref{fig:Cls})}, and therefore, we re-minimize the likelihood by varying $\mu_1$ and $\mu_{1,p}$. These HOD parameters act as an efficient proxy for bias and its redshift evolution. We obtain $\mu_1=13.05$ and $\mu_{1,p}=0.79$. These are used for the smoothed and perturbed curves in Fig.~\ref{fig:Nzs} and Fig.~\ref{fig:Cls} as well as Table~\ref{tab:lin}.  We have found that more aggressive smoothing produces a worse best-fit $\chi^2$ even after refitting. Finally the thick dotted yellow curve illustrates a realization of $\vN$ drawn from a multivariate Gaussian distribution with a mean given by the red line and a covariance given by ${\sf P}$ in Eq. \ref{eq:priorcov}.
      
      The angular power spectra estimated for the same set of three $N(z)$s are shown in Fig.~\ref{fig:Cls} using the same color scheme. The figure also shows the comparison between the exact prediction for the perturbed $N(z)$s (thick dotted yellow line) and the linear prediction around the smoothed distribution using the derivative matrix, ${\sf T}$ (dotted light-blue line). We find that the two are in reasonably good agreement. As demonstrated below using $\chi^2$ {(see Table~\ref{tab:lin})}, the agreement is sufficient given the measurement error, {which indicates that the Taylor expansion approximation is adequate for this study}.
      
      This may not be the case for future datasets with higher statistical power, although the expectation is that those data will be accompanied by larger spectroscopic samples \cite{2020arXiv200702631G} needed to reduce systematic photo-$z$ uncertainties. These improved datasets will naturally decrease the range over which the Taylor expansion needs to be sufficient. We also note that the agreement between exact prediction and linear model is noticeably worse for auto-correlations. As described in Appendix \ref{app:autos}, this is due to the fact that a single $N(z)$ enters the corresponding auto-correlation quadratically, making higher-order terms in the Taylor expansion more relevant than in the case of cross-correlations.

      In Table~\ref{tab:lin}, we show the $\chi^2$ values for all four curves with respect to the HSC data. Those are computed via
      \begin{equation}
          \chi^2 = ({\bf c}-{\bf t})^T \sC_X^{-1} ({\bf c}-{\bf t})
          \label{eq:chi2}
      \end{equation}
      where $\sC_X$ corresponds to either the orginal HSC covariance matrix, $\sC_c$, or the marginalized $N(z)$ covariance, $\sC_M$ (first and second column in Table~\ref{tab:lin}, respectively). Comparing the $\chi^2$ values for the original and smoothed $N(z)$ distributions, we see that they decrease for both the HSC covariance matrix and also for the marginalized one. As expected, perturbing the photometric distributions gets penalized, but the change in the $\chi^2$ value is much larger for the original HSC covariance ($\Delta\chi^2\simeq9$ compared to $\Delta\chi^2\simeq3$), which indicates that the marginalized covariance is more lenient in allowing small (and expected) $N(z)$ fluctuations. The difference between the last two lines suggests that the Taylor expansion works sufficiently well, as the differences in $\chi^2$ are negligible. 

      \begin{table}
        \begin{center}
          \begin{tabular}{l| c c }
            \hline\hline
            Model & HSC cov. & Marginalized $N(z)$ cov. \\ [0.5ex]
            \hline
            $\chi^2$, original $N(z)$, exact $C_\ell$ & 98.97 & 88.97 \\ 
            $\chi^2$, smoothed $N(z)$, exact $C_\ell$ & 91.33 & 86.44 \\
            $\chi^2$, perturbed $N(z)$, exact $C_\ell$ & 99.78 & 89.57 \\
            $\chi^2$, perturbed $N(z)$, $C_\ell(\bar{\vN}) + {\sf T} \Delta \vN$ & 100.62 & 89.64 \\
            \hline
            \hline
          \end{tabular}
        \end{center}
        \caption{$\chi^2$ values with respect to the HSC data of the four curves shown in Fig.~\ref{fig:Cls} using the original covariance used in the HSC analysis and the marginalized $N(z)$ covariance developed in this work (see Eq.~\ref{eq:chi2}). The total number of degrees of freedom is 94 and the effective degrees of freedom from the modified covariance matrix is 88.32 (see Eq.~\ref{eq:dof}). The perturbed $N(z)$ is penalized by only 3.1 units in $\chi^2$ compared to 8.4 for the original covariance matrix. The small differences between the third and fourth lines demonstrate that the Taylor expansion is sufficient to describe the allowed deviations in the $N(z)$.}\label{tab:lin}
      \end{table}

    \subsection{Impact of $N(z)$ uncertainties on final parameters}\label{ssec:hcs.mcmc}
      \begin{figure}
        \centering
        \begin{subfigure}{.5\textwidth}
          \centering
          \includegraphics[width=1.\linewidth]{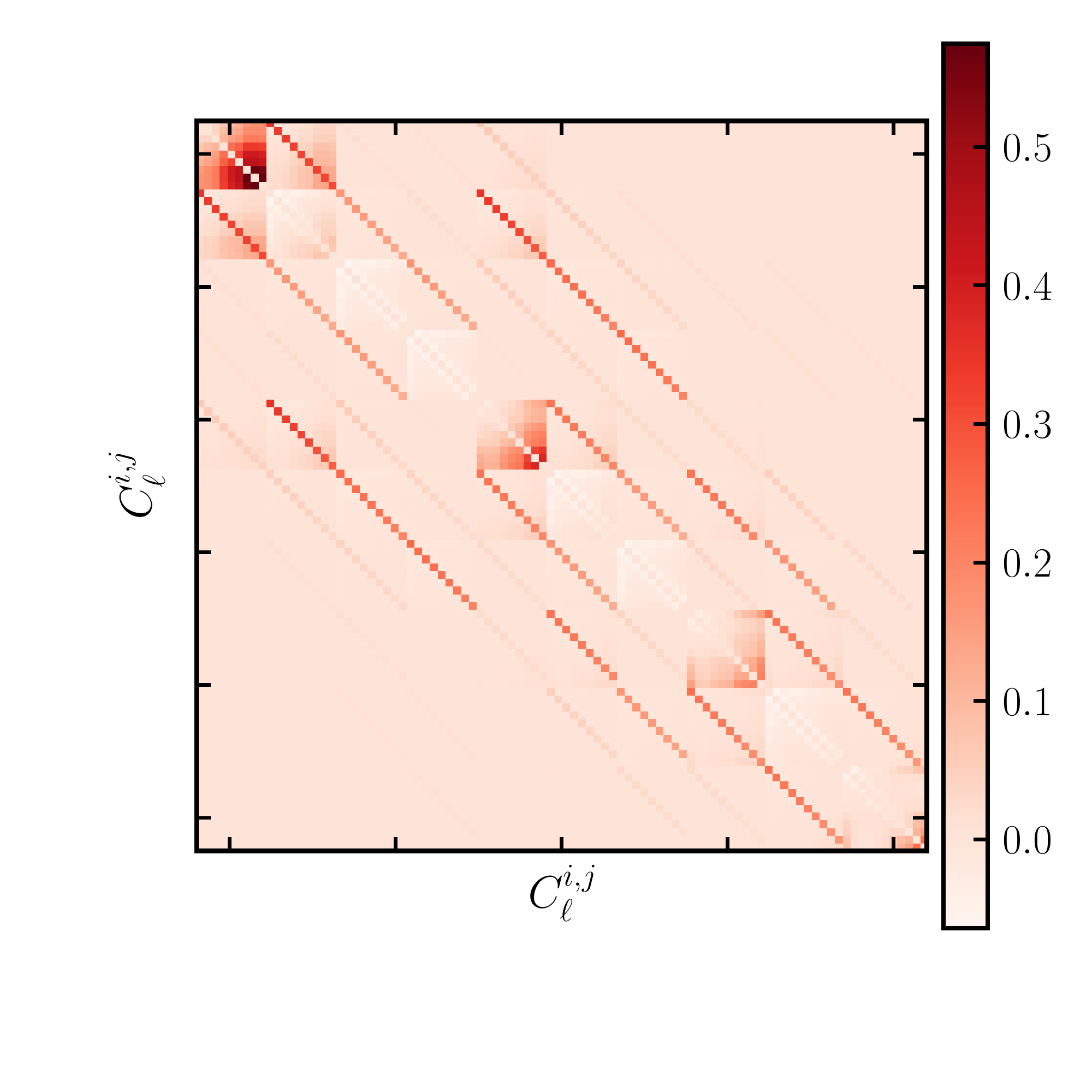}
        \end{subfigure}%
        \begin{subfigure}{.5\textwidth}
          \centering
          \includegraphics[width=1.\linewidth]{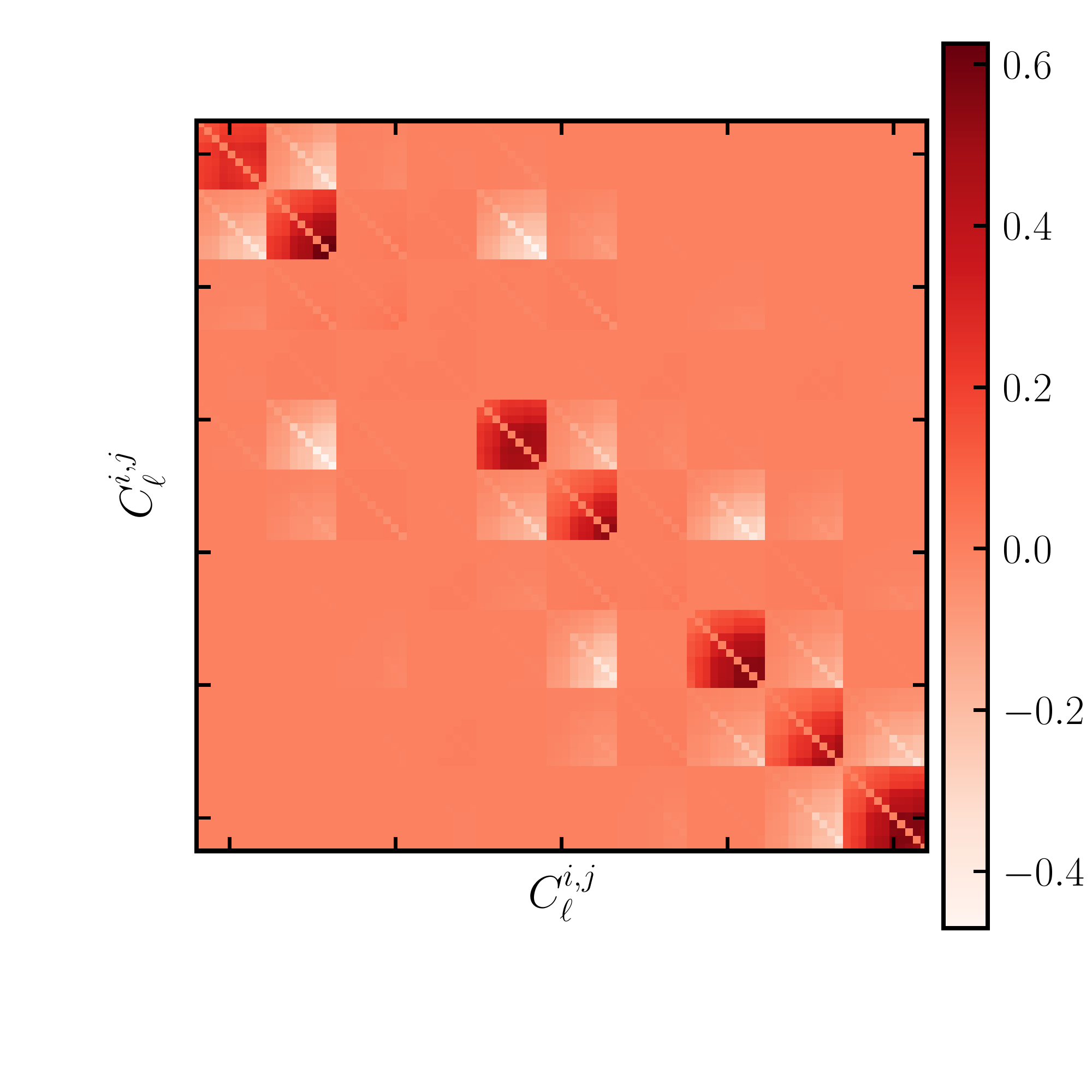}
        \end{subfigure}
        \caption{Correlation matrix of the original HSC dataset (\textit{left} panel) and difference between the original and the marginalized (this work) correlation matrices (\textit{right} panel). Most of the effect of this marginalization is manifested as a positive contribution to the original matrix near the diagonal (i.e. there is an increase in correlation for close redshift values). Note that we have subtracted the diagonal off the two matrices.}\label{fig:fid_marg_cov}
      \end{figure}
      \begin{figure}
        \centering  
        \includegraphics[width=1.\textwidth]{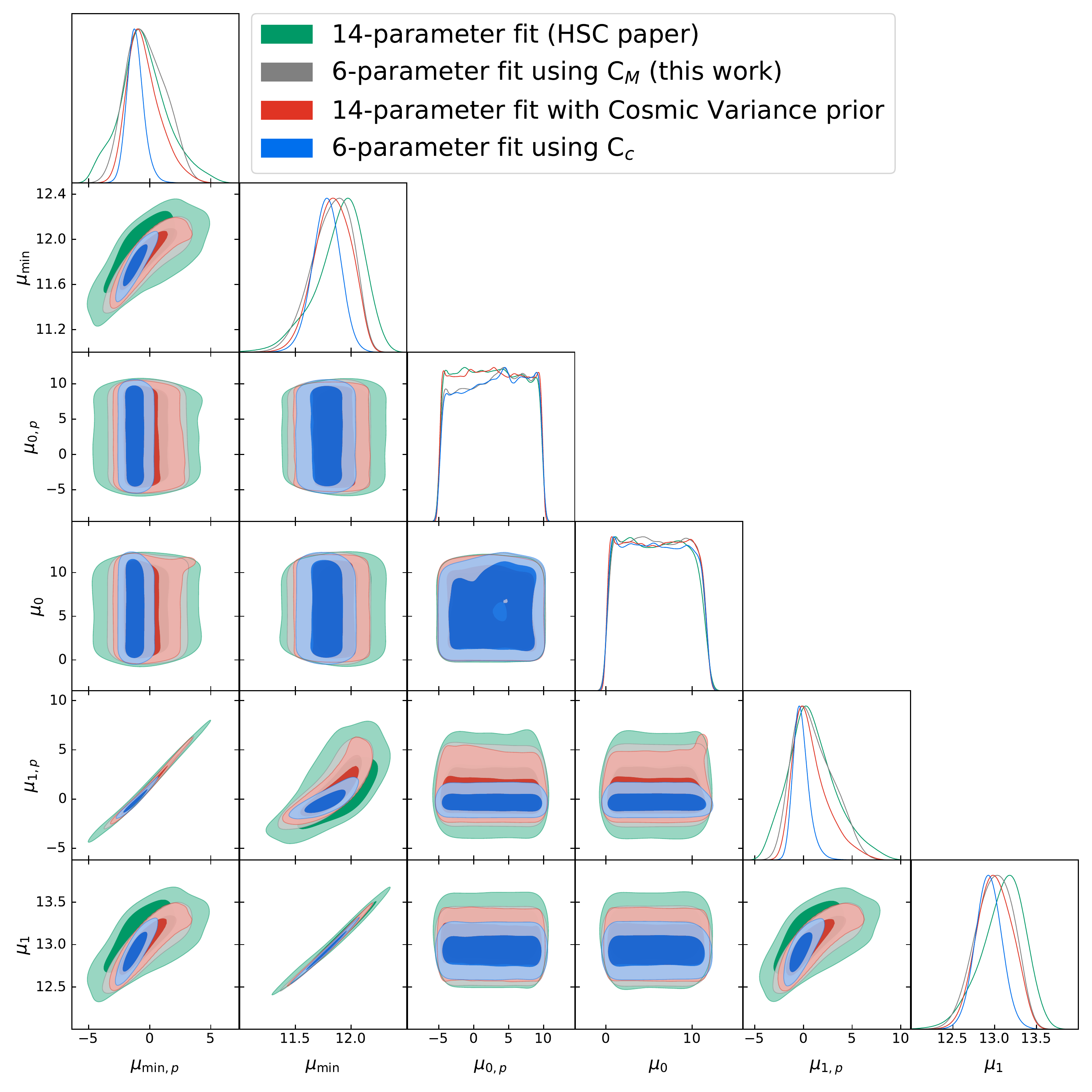}
        \caption{Triangle plot with constraints on the 6 HOD parameters obtained using the  covariance matrix from the full 14-parameter original HSC analysis (\textit{green contours}). The \textit{red contours} are obtained by running an MCMC chain with the original HSC covariance and a Gaussian prior on the 4 shift and 4 width parameters accounting for cosmic variance uncertainties. In \textit{gray}, we show the constraints on the 6 HOD parameters obtained using the covariance matrix, marginalized to account for the $N(z)$ uncertainty (i.e. without the 8 shift and width parameters). {The analytical marginalization recovers constraints similar to, but slightly broader than, those found with the shift-width parametrization. This indicates that, while this parametrization likely accounts for the bulk of the uncertainty in the $N(z)$, the analytical method automatically marginalizes over residual uncertainty modes without including any nuisance parameters. For context, the \textit{blue} contours show the constraints derived by fixing the shift and width parameters and using the original covariance matrix. The two HOD parameters $\mu_{0,p}$ and $\mu_{0}$ cannot be constrained with the current HSC data, so the contours recover our input priors (see \cite{1912.08209} for a detailed discussion)}.} \label{fig:triangle_fid_marg}
      \end{figure}
      In this section, we summarize the main results of this paper, showing in particular the effect of marginalization over the photometric uncertainties on the final parameter constraints. In our particular case, these are the 6 HOD parameters described in Section~\ref{ssec:hsc.theory}. The ${\sf T}$ matrix used to calculate the marginalized $N(z)$ covariance was computed assuming the fiducial HOD parameters listed in Eq. \ref{eq:par_fid}.

      In order to compute the $N(z)$ covariance ${\sf P}$ (see Section \ref{ssec:theory.prior}), we set the relevant, free parameters to
      \begin{equation}
        (A_{\rm smooth},\,A_{\rm noise},\,N_{\rm up},\,\Delta z_{\rm thr})=(1,\,4,\,3,\,0.06).
      \end{equation}
      The value of the smoothing amplitude $A_{\rm smooth}$, which produces only a mild effect on the $N(z)$ (see Fig. \ref{fig:Nzs}) was chosen so as to ensure that the resulting smoothed $N(z)$ was still able to describe the measured power spectra with a good $\chi^2$ value. We choose to set the additional noise due to photo-$z$ systematic uncertainties to be $A_{\rm noise}=4$ times larger than the variance in the different photometric redshift codes, so as to provide a conservative estimate of the expected variation in $N(z)$.
      
      In order to study the effect of marginalization (described in detail in Section~\ref{ssec:theory.nz_new}) on the covariance matrix, we visualize the original correlation matrix and its difference with the marginalized matrix in Fig.~\ref{fig:fid_marg_cov}. The main difference with the original matrix is a positive contribution to the off-diagonal entries for redshift samples in close proximity.
      
      \begin{table}
        \begin{center}
          \begin{tabular}{c | c c c c} 
            \hline\hline
            Parameter & HSC cov. [68\%, 95\%] & CV constraints [68\%, 95\%] & Marg. $N(z)$ [68\%, 95\%] \\ [0.5ex] 
            \hline
            $\chi^2/\nu$ & 87.49/80 & 88.29/80 & 88.54/82.32 \\ 
            $\mu_{{\rm min},p}$ & $-0.5^{+1.7}_{-2.0}$, $^{+4.0}_{-4.0}$ & $-0.44^{+0.96}_{-1.6}$, $^{+2.9}_{-2.4}$ & $-0.3^{+1.5}_{-1.8}$, $^{+3.2}_{-2.8}$ \\ [1ex]
            $\mu_{\rm min}$ & $11.90^{+0.26}_{-0.15}$, $^{+0.39}_{-0.46}$ & $11.83^{+0.19}_{-0.15}$, $^{+0.31}_{-0.31}$ & $11.83^{+0.21}_{-0.15}$, $^{+0.32}_{-0.35}$ \\ [1ex]
            $\mu_{0,p}$ & $2.4\pm 4.3$, $^{+7.2}_{-7.0}$ & $2.5\pm 4.3$, $^{+7.1}_{-7.1}$ & $2.7^{+6.7}_{-4.4}$, $^{+6.9}_{-7.1}$ \\ [1ex]
            $\mu_{0}$ & $5.7\pm 3.3$, $^{+5.5}_{-5.5}$ & $5.8\pm 3.4$, $^{+5.5}_{-5.6}$ & $5.8\pm 3.4$, $^{+5.5}_{-5.5}$ \\ [1ex]
            $\mu_{1,p}$ & $0.9^{+2.0}_{-2.8}$, $^{+5.1}_{-4.7}$ & $0.8^{+1.0}_{-2.2}$, $^{+4.0}_{-3.0}$ & $1.0^{+1.6}_{-2.6}$, $^{+4.1}_{-3.5}$ \\ [1ex]
            $\mu_{1}$ & $13.10^{+0.30}_{-0.21}$, $^{+0.48}_{-0.54}$ & $12.99\pm 0.24$, $^{+0.41}_{-0.39}$ & $13.00^{+0.25}_{-0.21}$, $^{+0.42}_{-0.44}$ \\ [1ex]
            \hline\hline
          \end{tabular}
        \end{center}
        \caption{Minimum $\chi^2$ values (divided by the degrees of freedom) reached in each of the three MCMC runs and [68\%, 95\%] constraints on the 6 HOD parameters in the following scenarios: 1) adopting the original HSC covariance matrix with 14 parameters (6 HOD + 8 $N(z)$ ones), 2) adopting the original HSC covariance matrix with Gaussian priors on the $N(z)$ parameters determined by their statistical distributions due to cosmic variance, and 3) using the marginalized $N(z)$ covariance matrix developed in this work to constrain the 6 HOD parameters. The degrees of freedom in the full 14-parameter case are ${\rm d.o.f.} = 94-14=80$, whereas for the marginalized covariance with a 6-parameter fit, the effective number of degrees of freedom is ${\rm d.o.f.} = 88.32-6=82.32$ (see Eq.~\ref{eq:dof}).}\label{tab:chi2_fid}
      \end{table}

      To compare the results of the automatic marginalization procedure proposed in this work with those of the original shift-width parametrization in terms of the final constraints on the 6 HOD parameters, we explore the posterior distribution of these parameters in both cases using the MCMC ensemble sampler ${\tt emcee}$ \cite{2013PASP..125..306F}. We test the convergence of the parameter chains via a simple Gelman-Rubin convergence diagnostic after analyzing the auto-correlation statistics.

      In Fig.~\ref{fig:triangle_fid_marg} and Table~\ref{tab:chi2_fid}, we illustrate the key  comparison between the original HSC analysis, marginalizing over the 8 shift-width parameters, and the marginalization scheme developed in this paper. The fiducial analysis plotted in green contours is the most conservative within the variable width and shift paradigm, since these parameters were given flat uninformative priors (i.e. the parameters were limited by how far the data allowed them to go, not by the prior). To directly compare the shift and width parametrization with our approach, we repeat the full 14-parameter fit, but impose a Gaussian prior on the shift-width parameters that allows them to vary consistently within the cosmic variance uncertainties in $N(z)$. These priors have zero mean and a standard deviation $\sigma(\Delta z_i)=0.008$, $\sigma(z_{w,i})=0.05$, and were obtained by computing the scatter in the mean and relative width of random redshift distributions drawn from a multivariate Gaussian with covariance given by ${\sf C}_{N,{\rm CV}}$ in Eq. \ref{eq:nz_cv}. The corresponding constraints are plotted in red. {For context, the blue contours show the constraints we infer when eliminating all shift and width parameters, assuming the redshift distribution is known perfectly.} Finally, in gray we plot constraints plotted using our approach. We find that using informative priors on shifts and widths shrinks the contours significantly and yields contours that are almost indistinguishable from those obtained by our approach. This explicitly confirms the intuition that shifts in the mean redshift and changes to the widths of the photo-$z$ distributions are in effect the most important contributions to degeneracies that decrease the sensitivity to cosmological parameters. {This plot suggests that accounting for the $N(z)$ uncertainties is important, and that although the shift and width parameters are able to account for them to an extent, our method is able to do this in a conscientious way with no extra parameters.} {We note that the HOD parameters $\mu_{0,p}$ and $\mu_{0}$ cannot be constrained with the current data, so the confidence intervals stated in this study reflect our input priors (see \cite{1912.08209} for a detailed discussion)}.

\begin{figure}[ht]
        \centering
        \includegraphics[width=1.\textwidth]{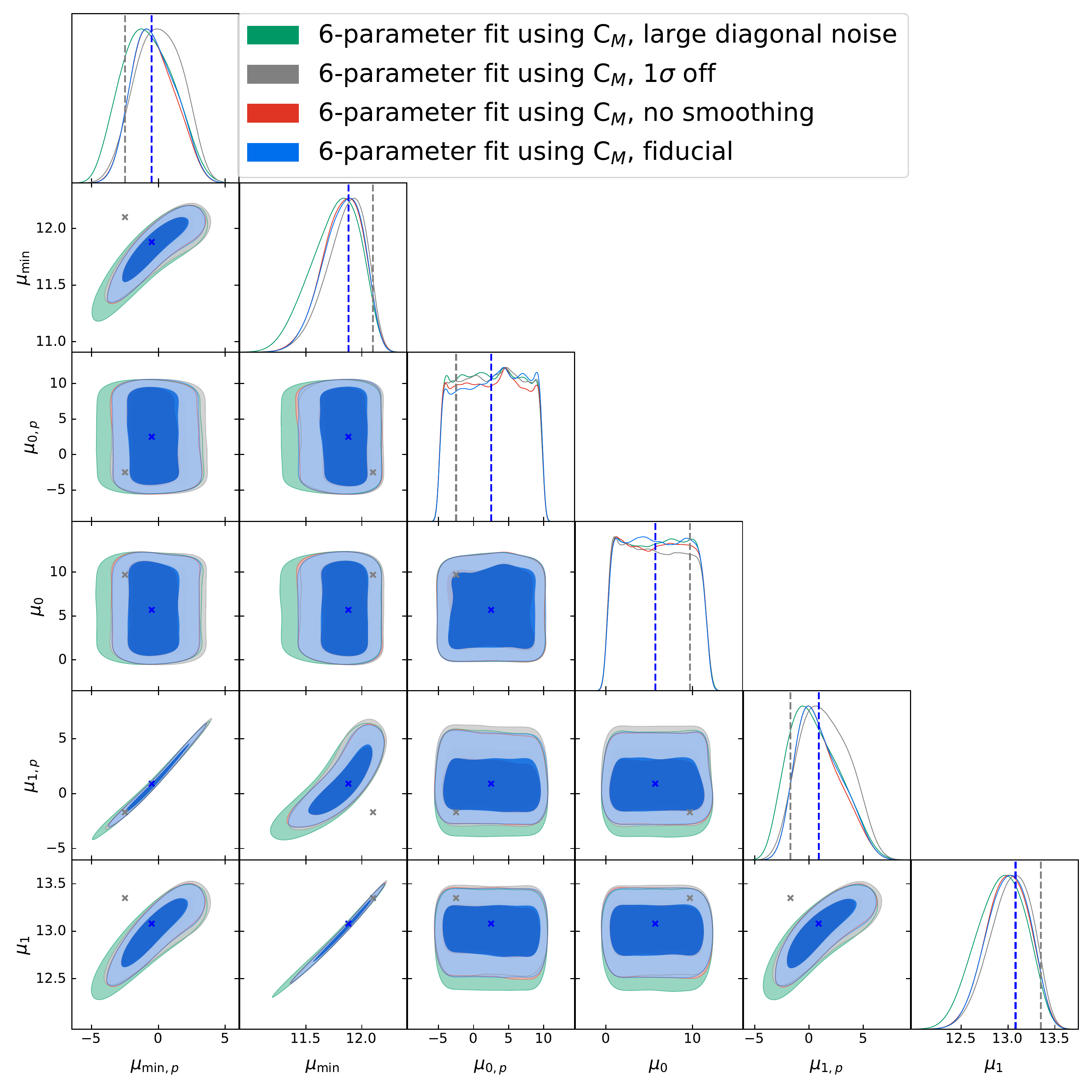}
        \caption{Triangle plot with constraints on the 6 HOD parameters obtained using different choices for the marginalized covariance matrix: in \textit{green}, we show the result where the diagonal noise is increased significantly ($A_{\rm noise} = 42$); in \textit{gray}, we show the case where the marginalization is performed assuming fiducial HOD-model values that are $1\sigma$ away (see Eq.~\ref{eq:par_t3}) from their posterior mean  values taken from Ref.~\cite{1912.08209}; in \textit{red}, we display the case where the smoothing has been removed ($A_{\rm smooth} = 0$); and finally the \textit{blue} contours show the fiducial case with $A_{\rm noise} = 4$ and $A_{\rm smooth} = 1$. The fiducial values of the HOD parameters used in this analysis are shown as \textit{dotted blue} lines in the 1D plots and blue crosses in the 2D ones (also listed in Section \ref{ssec:hsc.lin}), while those for the $1\sigma$-offset-test are shown as \textit{dotted gray} lines and gray crosses (see their values in Section \ref{ssec:hcs.mcmc}).} \label{fig:triangle_marg_tests}
      \end{figure}

      To check the robustness of our result, we perform the following tests:
      \begin{itemize}
        \item[\textbf{Test 1}:] We study the effect of the smoothing prior by turning it off, $A_{\rm smooth} = 0$.
        \item[\textbf{Test 2}:] We vary the choice for the diagonal photometric noise parameter by increasing it ten-fold, $A_{\rm noise} = 42$. 
        \item[\textbf{Test 3}:] We compute the ${\sf T}$ matrix using a different set of HOD parameters perturbed by $1\sigma$ relative to the fiducial ones (listed in Section \ref{ssec:hsc.lin}):
        \begin{equation}\label{eq:par_t3}
        (\mu_{\rm min},\mu_{{\rm min},p},\mu_0,\mu_{0,p},\mu_1,\mu_{1,p})=(12.10,-2.5,9.7,-2.5,13.35,-1.9).
      \end{equation}
      \end{itemize}
      We show the comparisons between these three tests and the fiducial marginalized $N(z)$ model in Fig. \ref{fig:triangle_marg_tests} and Table~\ref{tab:chi2_tests}. One can notice that the effect of the smoothing on the constraints is negligible, while increasing the diagonal noise weakens them. Neither of these tests affect the mean of the posterior distribution. On the other hand, offsetting the HOD parameters used in the ${\sf T}$ matrix computation leads to a slight shift and broadening of the contours. However, this is expected, since the covariance matrix in this case is significantly far from the favored model parameters. Thus, our use of a ${\sf T}$ matrix evaluated only once at the fiducial parameters is justified as long as the parameters chosen to compute it are within reasonable limits.

    \begin{table}
      \begin{center}
        \begin{tabular}{c | c c c c} 
          \hline\hline
          Parameter & Fiducial [68\%, 95\%] & Test 1 [68\%, 95\%] & Test 2 [68\%, 95\%] & Test 3 [68\%, 95\%] \\ [0.5ex] 
          \hline
          $\chi^2/\nu$ & 88.54/82.32 & 87.67/82.32 & 88.56/82.32 & 78.61/82.32 \\
          $\mu_{\rm min}$ & $11.83^{+0.21}_{-0.15}$, $^{+0.32}_{-0.36}$ & $11.85^{+0.22}_{-0.13}$, $^{+0.31}_{-0.37}$ & $11.83^{+0.21}_{-0.15}$, $^{+0.33}_{-0.36}$ & $11.76^{+0.26}_{-0.17}$, $^{+0.38}_{-0.43}$ \\ [1ex]
          $\mu_{0}$ & $5.8\pm 3.4$, $^{+5.5}_{-5.5}$ & $5.8^{+3.6}_{-5.2}$, $^{+5.6}_{-5.5}$ & $5.8\pm 3.4$, $^{+5.5}_{-5.5}$ & $5.9\pm 3.4$, $^{+5.5}_{-5.6}$ \\ [1ex]
          $\mu_{1}$ & $13.00^{+0.25}_{-0.21}$, $^{+0.42}_{-0.44}$ & $13.03^{+0.26}_{-0.20}$, $^{+0.41}_{-0.45}$ & $13.00^{+0.25}_{-0.22}$, $^{+0.42}_{-0.44}$ & $12.93^{+0.29}_{-0.24}$, $^{+0.47}_{-0.50}$ \\ [1ex]
          $\mu_{{\rm min},p}$ & $-0.3^{+1.5}_{-1.8}$, $^{+3.2}_{-2.8}$ & $0.0\pm 1.7$, $^{+3.2}_{-3.3}$ & $-0.3^{+1.4}_{-1.8}$, $^{+3.2}_{-2.8}$ & $-0.8^{+1.8}_{-2.1}$, $^{+3.7}_{-3.5}$ \\ [1ex]
          $\mu_{0,p}$ & $2.7^{+6.7}_{-4.4}$, $^{+6.9}_{-7.1}$ & $2.5\pm 4.3$, $^{+7.1}_{-7.1}$ & $2.5\pm 4.3$, $^{+7.1}_{-7.1}$ & $2.5\pm 4.3$, $^{+7.1}_{-7.1}$ \\ [1ex]
          $\mu_{1,p}$ & $1.0^{+1.6}_{-2.6}$, $^{+4.1}_{-3.5}$ & $1.5^{+2.1}_{-2.6}$, $^{+4.3}_{-3.9}$ & $0.98^{+1.5}_{-2.5}$, $^{+4.2}_{-3.5}$ & $0.6^{+1.9}_{-2.9}$, $^{+4.6}_{-4.0}$ \\ [1ex]
          \hline
          \hline
        \end{tabular}
      \end{center}
      \caption{Minimum $\chi^2$ values (divided by the degrees of freedom) attained in each of the four MCMC runs and [68\%, 95\%] constraints on the 6 HOD parameters. The ``fiducial'' column displays the results for the marginalized $N(z)$ covariance matrix adopting the fiducial parameter choices described in the text, while the three tests refer to the marginalized $N(z)$ covariance matrix obtained with no smoothing, large diagonal noise and $1\sigma$-offset HOD parameters, respectively. The effective number of degrees of freedom for the marginalized covariance with a 6-parameter fit is ${\rm d.o.f.} = 88.32-6=82.32$ (see Eq.~\ref{eq:dof}).}\label{tab:chi2_tests}
    \end{table}

  \section{Conclusions}\label{sec:conclusions}
    In this paper, we have presented a new method to marginalize over uncertainties in the redshift distribution in tomographic large-scale structure analyses. This method consists of modifying the data covariance matrix by commensurately increasing the variance of the modes that are most sensitive to these uncertainties. This is akin to mode-deprojection methods used in other areas of cosmological data analysis, such as power spectrum estimation or component separation.

    We have demonstrated the performance of this method by applying it to the analysis of photometric galaxy clustering data from HSC DR1, comparing it with previous approaches. The method performs as expected, propagating the uncertainties in $N(z)$ without including any additional nuisance parameters. In general this method is applicable for any 3$\times$2-point analysis combining galaxy clustering and cosmic shear. Since galaxy clustering is considerably more sensitive to redshift distribution uncertainties than cosmic shear (given the radially cumulative nature of the latter), the method will likely perform even better in that case.
    
    This method relies on three approximations. The first approximation is that a first-order expansion of the theory data vector with respect to a change in redshift distribution is sufficient over the range of interest. The range of interest is given by the ${\sf P}$ matrix, which determines how far from the best guess we allow our $N(z)$s to wander. The second one is that the derivative of the theory prediction with respect to the redshift distribution amplitudes can be approximated to be constant over the parameter space of interest. We have explicitly shown that the first approximation is good enough for existing data from the COSMOS 30-band sample. As our measurements improve, the $N(z)$ uncertainties will shrink, making this approximation more reliable. We have also shown that the second approximation is valid by re-evaluating our parameter constraints for the first derivative matrix (${\sf T}$, see Eq.~\ref{eq:Tmat}) calculated away from the best-fit parameters. Note that the main reason to avoid recomputing ${\sf T}$ at every point in parameter space is that its calculation can take significantly longer than the calculation of the theory prediction ${\bf t}$. This could however be sped up by incorporating the calculation of ${\sf T}$ as a product of the Limber integrator used to compute ${\bf t}$. Alternatively, this approximation could be improved by expanding ${\sf T}$ itself to linear order in the cosmological and astrophysical parameters:
    \begin{equation}
      {\sf T} ({{\bf q},\bar{\vN}}) = {\sf T}_{\rm fid} + \left. \frac{\partial \sf T}{\partial {\bf q}}\right|_{{\bf q_{\rm fid}}} \left({\bf q} - {\bf q_{\rm fid}}\right). \label{eq:Tmatd}
    \end{equation}
    We leave this study for future work. In short, these two approximations are sufficient for the present analysis and can be improved further, if necessary.

    The third approximation used here is the assumption that the redshift distribution uncertainties follow a multivariate Gaussian distribution. This is perhaps not true in detail for direct calibration methods, where Poisson or Dirichlet distributions may be more appropriate in certain regimes \cite{2019arXiv191007127A,2004.09542}. On the other hand, the $N(z)$ uncertainties for clustering redshifts measurements \cite{2008ApJ...684...88N} would likely be well described by Gaussian statistics. We argue that, regardless of the specifics of the underlying distribution, the Gaussian assumption allows us to quantify the allowed statistical variations in the $N(z)$ (potentially overestimating them), and to easily propagate them into the relevant modes of the theory data vector, leading to reliable parameter constraints marginalized over these uncertainties. A detailed validation of this assumption is also left for future work. With the development of new photo-$z$ methods, the characterization of these distributions will most certainly become an integral part of photometric redshift and $N(z)$ estimation processes \cite{2004.09542}.

    The central problem in implementing this method is the determination of the covariance matrix of the $N(z)$ uncertainties, $\sC_N$, as well as any physical priors on e.g. the smoothness of the underlying distribution. In this paper, we have considered two different contributions to $N(z)$ uncertainties: first, contributions from sample variance in the COSMOS-30 field; and secondly, we have included an additional diagonal noise component given by the scatter between the redshift distributions estimated by stacking of individual source pdfs from four different photo-$z$ codes. To account for common systematic uncertainties between these codes, we have multiplied the resulting noise by a ``safety'' factor. We also implemented a ``smoothness'' prior following the intuition that the true $N(z)$ must be smooth and hence it is unnecessary to marginalize over non-physical $N(z)$s. As we have shown, for the mild prior used here, the effect is largely negligible.

    With growing datasets and shrinking statistical uncertainties, the analysis of current and future photometric surveys will require a careful propagation of uncertainties in the redshift distributions of the different samples considered. A shift from ad-hoc parametrizations in terms of e.g. variations in the mean and width of the distributions, to more principled methods that account for all possible modes of variation in the $N(z)$, will be necessary in order to obtain reliable constraints on cosmological parameters. The method presented here provides a computationally efficient and accurate way to move in that direction.

\section*{Acknowledgements}
  {We would like to thank Dragan Huterer and Alex Hall for their insightful comments, which sparked fruitful discussions}. DA acknowledges support from Science and Technology Facilities Council through an Ernest Rutherford Fellowship, grant reference ST/P004474/1. AS is supported by the U.S. Department of Energy, Office of Science, Office of High Energy Physics, under Contract No. DE-SC0012704. {AN acknowledges support from National Science Foundation Grant No. 1814971. This publication arises from research funded by the John Fell Oxford University Press Research Fund.}

  The Hyper Suprime-Cam (HSC) collaboration includes the astronomical communities of Japan and Taiwan, and Princeton University. The HSC instrumentation and software were developed by the National Astronomical Observatory of Japan (NAOJ), the Kavli Institute for the Physics and Mathematics of the Universe (Kavli IPMU), the University of Tokyo, the High Energy Accelerator Research Organization (KEK), the Academia Sinica Institute for Astronomy and Astrophysics in Taiwan (ASIAA), and Princeton University. Funding was contributed by the FIRST program from Japanese Cabinet Office, the Ministry of Education, Culture, Sports, Science and Technology (MEXT), the Japan Society for the Promotion of Science (JSPS), Japan Science and Technology Agency (JST), the Toray Science Foundation, NAOJ, Kavli IPMU, KEK, ASIAA, and Princeton University. 

  This paper is based on data collected at the Subaru Telescope and retrieved from the HSC data archive system, which is operated by Subaru Telescope and Astronomy Data Center at National Astronomical Observatory of Japan. Data analysis was in part carried out with the cooperation of Center for Computational Astrophysics, National Astronomical Observatory of Japan.

\appendix
  \section{Precision of the linear expansion for auto- and cross-correlations}\label{app:autos}
    In order to quantify the validity of the first-order expansion of power spectra in the $N(z)$ uncertainties, let us start by discretizing the redshift distribution as in Eq. \ref{eq:n_discrete}. Let $\vN_i\equiv (N_i^1,...,N_i^{n_z})$ be the vector of all $N(z)$ amplitudes for the $i$-th redshift bin. For simplicity we will assume that the basis functions $\phi_\alpha$ are normalized to unit area, so that the integral of the redshift distribution is simply $\sum_\alpha N^\alpha_i\equiv\vN_i^T\,{\bf 1}$, where ${\bf 1}=(1,1,...,1)$.

    With this notation, the cross-correlation between bins $i$ and $j$ (Eq. \ref{eq:cell_gg_limber}) is given by:
    \begin{equation}
      C^{ij}_\ell=\frac{{\vN}_i^T{\sf D}_\ell{\vN}_j}{({\vN}_i^T{\bf 1})({\vN}_j^T{\bf 1})},
    \end{equation}
    where ${\sf D}$ is the matrix of power spectra associated with the basis functions:
    \begin{equation}
      D^{\alpha\beta}_\ell\equiv \int dz\frac{H}{\chi^2}\phi_\alpha(z)\phi_\beta(z)\,P_{gg}\left(z,\frac{\ell+1/2}{\chi}\right).
    \end{equation}
    The gradient of $C^{ij}_\ell$ with respect to $\vN_k$ can be calculated analytically:
    \begin{equation}
      \frac{\partial C^{ij}_\ell}{\partial \vN_k}=\left(-\frac{C_\ell^{ij}}{\vN_i^T{\bf 1}}{\bf 1}+\frac{\vN_j^T{\sf D}_\ell}{({\vN}_i^T{\bf 1})({\vN}_j^T{\bf 1})}\right)\delta_{ik} + (i\longrightarrow j)
    \end{equation}

    Let us now consider a perturbed $N_i(z)$ with coefficients given by $\tilde{\vN}_i$. For simplicity, let us limit our discussion to perturbations that do not change the area under the $N(z)$, which we will further set to unity ($\vN_i^T{\bf 1}=\tilde{\vN}_i^T{\bf 1}=1$). The resulting perturbed auto-correlation of the $i$-th bin and its cross-correlation with the $j$-th bin are then given by
    \begin{equation}
      \tilde{C}^{ii}_\ell=\tilde{\vN}_i^T{\sf D}_\ell\tilde{\vN}_i,
      \hspace{12pt}
      \tilde{C}^{ij}_\ell=\tilde{\vN}_i^T{\sf D}_\ell{\vN}_j,
    \end{equation}
    and their expansion to first order in $\Delta\vN_i\equiv\tilde{\vN}_i-\vN_i$ (labelled $\grave{C}^{ij}_\ell$ here) is
    \begin{equation}
      \grave{C}^{ii}_\ell=\tilde{C}^{ii}_\ell-\Delta\vN{\sf D}_\ell\Delta\vN,
      \hspace{12pt}
      \grave{C}^{ij}_\ell=\tilde{C}^{ij}_\ell.
    \end{equation}
    Therefore, in this particular case, the linear expansion in one of the $N(z)$s is exact for cross-correlations, but leads to a second-order residual in auto-correlations. Of course, when we vary multiple redshift distributions at the same time, the cross-correlation will receive a quadratic term of the kind $\Delta \vN_i {\sf D}_\ell \Delta \vN_j$, which however, for normally distributed deviations averages to zero.
    
    To build a better intuition, let us specify the result above to the case of a {\sl shift} of the form:
    \begin{equation}
      \tilde{N}_i^\alpha=(1-x)N_i^\alpha+x\,N_i^{\alpha+1}.
    \end{equation}
    For $x=1$, $\tilde{N}_i^\alpha=N_o^{\alpha+1}$, and the perturbation becomes a shift of the power spectrum by an increment $\Delta z$ given by the width of the top-hat basis functions $\phi_\alpha$. For simplicity let us consider flat, noise-like power spectra (${\sf D}_\ell=D\mathbb{1}$), in which case the perturbed auto-correlation is
    \begin{equation}
      \tilde{C}^{ii}_\ell=C_\ell^{ii}+2x(x-1)D\sum_\alpha N^\alpha_i(N^\alpha_i-N^{\alpha+1}_i),
    \end{equation}
    which results in the original power spectrum for $x=0$ and 1, but not for intermediate values. The first-order expansion, on the other hand, is given by the linear term in $x$ in the previous equation:
    \begin{equation}
      \grave{C}_\ell^{ii}=C_\ell^{ii}-2xD\sum_\alpha N^\alpha_i(N^\alpha_i-N^{\alpha+1}_i),
    \end{equation}
    which is monotonically decreasing in $x$.
    
    Thus we see that, due to the way in which $N(z)$ enters auto-correlations, the first- and second-order terms in the Taylor series take opposite signs, and the first-order expansion becomes less accurate than in the case of cross-correlations.

\bibliography{bibliography}

\end{document}